\documentclass[aps,prb,longbibliography,superscriptaddress,preprint]{revtex4-1}
\usepackage{mathrsfs}
\usepackage{amsmath}
\usepackage{amsfonts}
\usepackage{amssymb}
\usepackage{amsthm}
\usepackage{graphicx}
\usepackage{color}
\usepackage{hyperref}
\usepackage{bm}
\usepackage[caption=false]{subfig}
\usepackage{verbatim}

\newcommand{\ket}[1]{\left| {#1} \right\rangle}
\newcommand{\keteig}[1]{\left| {#1} \right\rangle_{\theta,\phi}}
\newcommand{\keteigin}[1]{\left| {#1} \right\rangle_{0,0}}

\begin{document}
\title{Antiferromagnetic magnon pseudospin: Dynamics and diffusive transport}
\author{Akashdeep Kamra}
\email{akashdeep.kamra@ntnu.no} 
\affiliation{Center for Quantum Spintronics, Department of Physics, Norwegian University of Science and Technology, NO-7491 Trondheim, Norway}
\author{Tobias Wimmer}
\affiliation{Walther-Mei{\ss}ner-Institut, Bayerische Akademie der Wissenschaften, 85748 Garching, Germany}
\affiliation{Physik-Department, Technische Universit\"{a}t M\"{u}nchen, 85748 Garching, Germany}
\author{Hans Huebl}
\affiliation{Walther-Mei{\ss}ner-Institut, Bayerische Akademie der Wissenschaften, 85748 Garching, Germany}
\affiliation{Physik-Department, Technische Universit\"{a}t M\"{u}nchen, 85748 Garching, Germany}
\affiliation{Munich Center for Quantum Science and Technology (MCQST), Schellingstr. 4, D-80799 M\"{u}nchen, Germany}
\author{Matthias Althammer}
\affiliation{Walther-Mei{\ss}ner-Institut, Bayerische Akademie der Wissenschaften, 85748 Garching, Germany}
\affiliation{Physik-Department, Technische Universit\"{a}t M\"{u}nchen, 85748 Garching, Germany}

\begin{abstract}
We formulate a theoretical description of antiferromagnetic magnons and their transport in terms of an associated pseudospin. The need and strength of this formulation emerges from the antiferromagnetic eigenmodes being formed from superpositions of spin-up and -down magnons, depending on the material anisotropies. Consequently, a description analogous to that of spin-1/2 electrons is demonstrated while accounting for the bosonic nature of the antiferromagnetic eigenmodes. Introducing the concepts of a pseudospin chemical potential together with a pseudofield and relating magnon spin to pseudospin allows a consistent description of diffusive spin transport in antiferromagnetic insulators with any given anisotropies and interactions. Employing the formalism developed, we elucidate the general features of recent non-local spin transport experiments in antiferromagnetic insulators hosting magnons with different polarisations. The pseudospin formalism developed herein is valid for any pair of coupled bosons and is likely to be useful in other systems comprising interacting bosonic modes. 
\end{abstract}


\maketitle

\section{Introduction}\label{sec:intro}

The bosonic excitations of ordered magnets - magnons - have become the active ingredient in an emerging paradigm for spin information transport and processing via magnetic insulators~\cite{Bauer2012,Chumak2015,KajiwaraNL,Baltz2018,HillebrandsMag,KathrinLogik,UchidaSSE,Althammer2018,Mook2014,Mook2017,Nakata2017,Onose2010,OhnumaSSE,Flebus2016}. This has been enabled in part by accomplishing electronic injection and detection of magnon spin by an adjacent normal metal with spin-orbit coupling~\cite{HirschSHE,SaitohISHE,Sinova2015,Valenzuela}. Magnons in a uniformly ordered ferromagnet~\footnote{Here, we use the term ``ferromagnet'' in a general sense to include ferrimagnets, such as yttrium iron garnet. Many of the experiments have been carried out on the latter material, while the theoretical models often treat it as a ferromagnet, for simplicity.} carry spin in a fixed direction and were first exploited for such non-local spin transport\cite{CornelissenMMR,SchlitzMMR,Wimmer2019,CornelissenTheory,ZhangMMR1,Li2016}. In this configuration, injection and detection of magnonic spin is accomplished using two spatially separated heavy metal electrodes via spin Hall effect~\cite{HirschSHE,SaitohISHE,Sinova2015,Valenzuela}. 

Subsequently, non-local spin transport in easy-axis antiferromagnetic insulators (AFIs) was demonstrated in similar devices~\cite{Klaui2018,Klaui2020}. These AFIs host spin-up and -down magnons as the eigenmodes such that an injection of spin along the up direction is achieved by inducing an excess of spin-up magnons~\cite{Shen2019,Troncoso2020}. Depending on the anisotropy landscape, the eigenmodes in AFIs can have a variable spin~\cite{Kamra2017,Rezende2019}, including zero spin for certain configurations such as an easy-plane anisotropy, raising the question whether such AFIs can mediate non-local spin transport. Recent experimental observations~\cite{Klaui_2020,Han2020,Wimmer2020} answer this question in the affirmative and highlight further nontrivial phenomena such as a modulation, including reversal, of the excitation spin interpreted in terms of an antiferromagnetic magnon Hanle effect~\cite{Wimmer2020}. The present manuscript develops an understanding of such non-local spin transport studies in AFIs with arbitrary anisotropies and eigenmodes.

The feature that makes bipartite AFIs unique with respect to ferromagnets in the present context is that AFI eigenmodes at a given wavevector occur in pairs and thus enable linear combinations or superpositions~\cite{Cheng2016,Kamra2017,Shen2020}. This is reminiscent of a two-level system and inspires associating a pseudospin with the antiferromagnetic magnons~\cite{Cheng2016,Daniels2018,Kawano2019,Shen2020,Wimmer2020}. Such an analogy has previously been discussed within the Landau-Lifshitz framework for describing the AFI excitations~\cite{Cheng2016,Daniels2018}. Considering a quantum field theoretic treatment, AFI magnons with an associated pseudospin can also be compared to itinerant electrons along with their spin. The corresponding mathematical analogy has also been invoked in predicting emergent spin-orbit coupling effects with AFI magnons~\cite{Kawano2019,Shen2020,Kawano2019B}, including topological states~\cite{Hasan2010,Tokura2019,Mook2014,Kawano2019,Liu2020}. Complementary to this similarity with electrons that enables a comparison of their eigenmodes, crucial differences arise from AFI magnons being bosonic excitations as will be discussed here.

In the present manuscript, we develop a quantum field theoretic pseudospin description of AFI eigenexcitations and their nonequilibrium states demonstrating it to be especially useful in understanding non-local magnonic spin transport. While sharing similarities with various two level systems or excitations with two states, antiferromagnetic magnons are found to provide a unique embodiment of pseudospin due to their non-conserved bosonic nature. In this way, the similarities and differences with respect to the case of spin-1/2 electrons are recognized and a consistent theory for spin transport in AFIs with arbitrary anisotropies is developed. This comparison allows us to gain physical insights based on the existing understanding of electronic spin transport~\cite{Fabian2007,Wu2010} while adequately accounting for the bosonic features of AFI magnons. This also enables an intuitive understanding of recent non-local magnon transport experiments in AFIs~\cite{Klaui2018,Han2020,Klaui_2020,Wimmer2020,Ross2020} and provides a simple framework for further predictions.
 
\section{Overview}\label{sec:overview}
In the present section, we provide a qualitative discussion of the physics reported herein and an overview of the results discussed in subsequent sections. For readers not interested in mathematical details, the present section strives to sum up the main messages and should suffice for a physical understanding. While we introduce the pseudospin description in a broader context of understanding coherently coupled bosonic modes, the discussion and assumptions are often biased by the subsequent goal of addressing diffusive magnonic transport in AFIs. The pseudospin description of AFI modes may take inspiration from photons~\cite{Lan2017,Han2020} or electrons~\cite{Kawano2019,Shen2020}, both of them being two-state excitations. In the present analysis, we largely compare AFI magnons with electrons. This is, in part, because AFI magnons scatter strongly~\cite{Flebus2019} and manifest diffusive transport~\cite{Shen2019,Troncoso2020} for a wide range of physical parameters, similar to itinerant electrons~\cite{Fabian2007}. Furthermore, while our magnon based description of an ordered AFI is, strictly speaking, a low temperature approximation, it has been found to work well even at high temperatures~\cite{Keffer1961,Bloch1962}.

Let us first outline some key assumptions made in our analysis. We assume a N\'eel ordered AFI and capitalize on the typical hierarchy of energy scales, i.e. exchange interaction is assumed to be much stronger than all other energy contributions. This enables a perturbative treatment of anisotropies and other non-universal, material dependent interactions. Further, the ground state is assumed to have the two sublattice magnetizations antiparallel and oriented collinear with the $z$-axis. Therefore, spin-up and -down AFI magnons carrying unit spin parallel to the N\'eel vector ($z$-axis) constitute our natural basis. Spin-nonconserving interactions treated as perturbation couple the basis modes and enable the formation of their superpositions~\cite{Kamra2017,Cheng2016,Kamra2019,Rezende2019,Liensberger2019}. This brings us to the first key difference that AFI magnon pseudospin bears with respect to electronic spin. The choice of spin quantization axis for spin-up and -down electronic states is largely arbitrary and a matter of convenience. Therefore the basis for describing electron spin can be chosen with respect to any convenient axis. For AFI magnons, the N\'eel vector fixes this direction and the corresponding spin-up and -down magnons constitute a preferred natural basis. As a corollary, the magnons may carry spin only along the equilibrium N\'eel vector direction, chosen to be $z$-axis here. This breaking of symmetry in the pseudospin space is intricately related to the symmetry-breaking associated with emergence of N\'eel order in the ground state.

\begin{figure}[tb]
	\begin{center}
		\includegraphics[width=85mm]{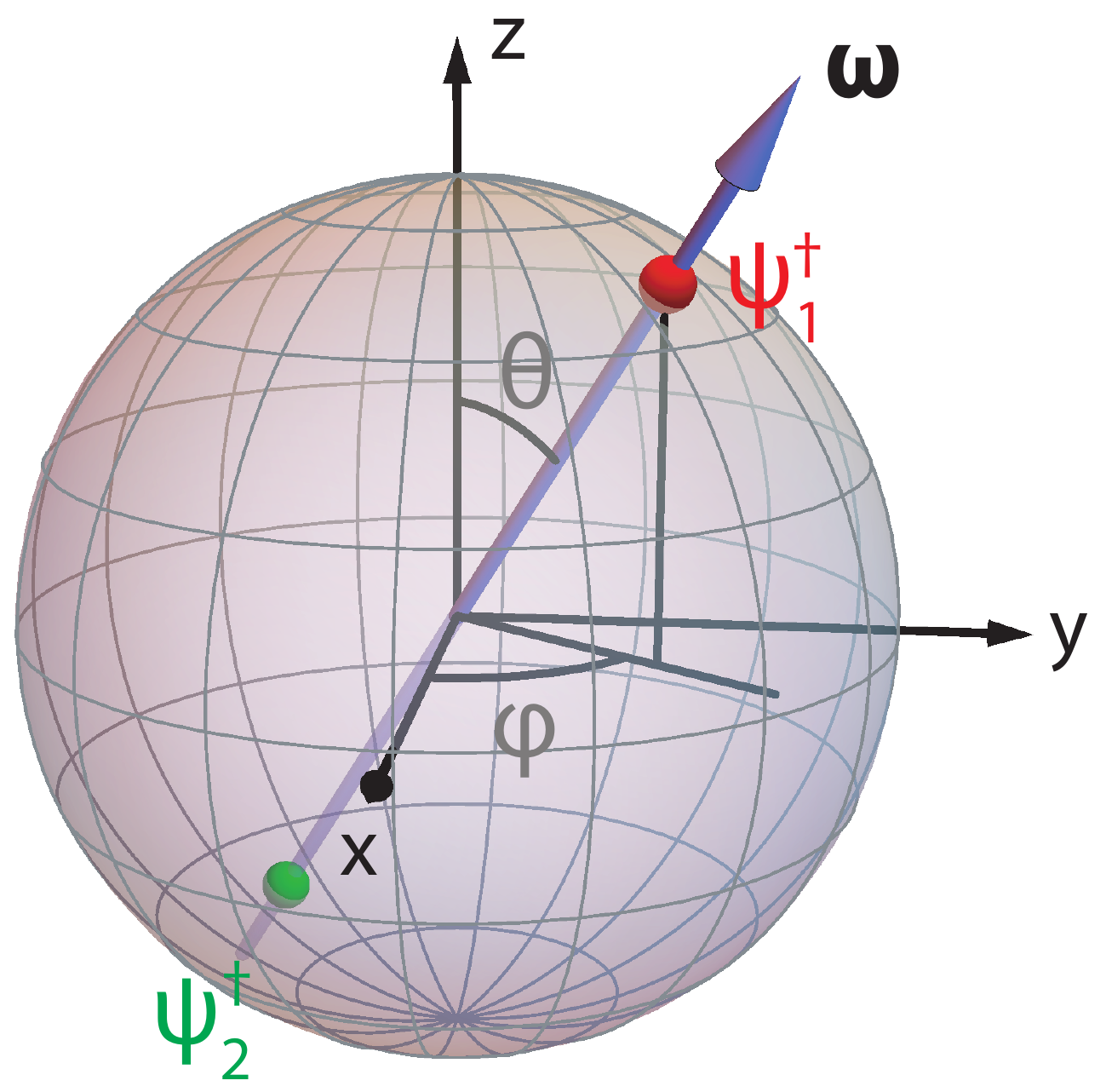}
		\caption{Schematic depiction of the two antiferromagnetic eigenmodes. The pseudofield vector [see Eq.~\eqref{eq:hampseudo}] depicted as a blue arrow intersects the Bloch sphere with unit radius at red and green points. These respectively represent the lower and higher energy magnonic eigenmodes [see Eq.~\eqref{eq:eigmodes}]. As discussed in the text, the depicted sphere is in the creation operator space.}
		\label{fig:BS}
	\end{center}
\end{figure}

In sections \ref{sec:eigenmodes} - \ref{sec:dynamics}, we analyze and develop the pseudospin description of two coupled bosonic modes. For concreteness, we consider the latter to be spin-up and -down AFI magnons disregarding their wavevector index. This allows us to introduce the pseudospin operator in terms of the bosonic modes' ladder operators. We show that the Hamiltonian for the coupled modes may be expressed as a dot product between the pseudospin operator and a fictitious pseudofield vector defined in terms of the coupling. The eigenmodes that result from a finite coupling may thus be represented by antiparallel unit vectors aligned with the pseudofield (Fig.~\ref{fig:BS}). For pseudofield pointing along $\hat{\pmb{z}}$, the eigenmodes are spin-up and -down magnons, corresponding to modes with circular precession of the N\'eel vector within the Landau-Lifshitz description. For pseudofield pointing in the $x$-$y$ plane, the eigenmodes are comprised by equal superpositions of the spin-up and -down magnons thereby bearing zero spin, and correspond to linear oscillation of the N\'eel vector. The actual eigenmode spin is proportional to the $z$ component of the pseudospin.

The measurable spin, however, is determined by both the eigenmodes and their occupation. Thus, in section \ref{sec:chemical}, we introduce a vector pseudospin chemical potential in order to capture the nature of eigenmodes and their degree of occupation in certain nonequilibrium situations. This also allows us to conveniently address pseudospin dynamics as discussed in section \ref{sec:dynamics}. We show that the pseudospin and its vector chemical potential precess about the pseudofield, analogous to the case of electron spin precessing about an applied magnetic field~\footnote{As per our chosen convention, the sense of precession for magnon pseudospin about pseudofield is opposite to that of electron spin about a magnetic field due to the negative gyromagnetic ratio of the latter.}. For AFI excitations, this pseudospin precession implies a transmutation between the various kinds of magnons with different spins or polarizations (circular and linear).

In section \ref{sec:diffusion}, we employ the pseudospin and pseudofield concepts to obtain a description of diffusive magnon transport in AFIs. We derive a semi-phenomenological diffusion equation for the total pseudospin density and its chemical potential, where the pseudofield averaged over the occupied modes is shown to induce coherent pseudospin precession dynamics. The spin and pseudospin decay are introduced via phenomenological anisotropic relaxation times. 

Solving the diffusion equation thus derived, in section \ref{sec:examples}, we investigate non-local magnon spin transport in an AFI. Here, we consider a general model for the AFI that continuously captures situations with spin-1 (circularly polarized) to spin-zero (linearly polarized) magnons as the eigenmodes. We find that, for the case of spin-zero eigenmodes, the non-local magnon spin signal manifests oscillations as a function of injector-detector distance and pseudofield magnitude caused by a pseudospin precession, consistent with the recently reported AFI magnon Hanle effect~\cite{Wimmer2020}. For spin-1 eigenmodes, we find the usual diffusive propagation of spin mediated by the magnons. For intermediate cases, we find an oscillating Hanle contribution that decays on shorter lengths and a positive offset~\cite{Wimmer2020} contributed by a longer range transport mediated by the finite spin of the eigenmode. We conclude with a summary in section \ref{sec:summary}. A perturbative derivation of the pseudofield for an example AFI model relevant to our discussion in section \ref{sec:examples} has been demonstrated in appendix \ref{sec:MH}.

\section{Pseudospin, pseudofield, eigenmodes, and magnonic spin}\label{sec:eigenmodes}
In the present section, we analyze the full range of eigenmodes admitted by two coupled bosonic modes and relate them to a spin-1/2 two-level system by describing the former in terms of a pseudospin operator and a fictitious pseudofield. For concreteness, we may consider the two bosonic modes to be spin-up and -down AFI magnons with a fixed wavevector and carrying a spin $\pm1$ along the $z$-axis.

\subsection{Two coupled modes}
We consider two coherently coupled bosonic modes described by the Hamiltonian $\tilde{H}$:
\begin{eqnarray}
\tilde{H}  & = & \omega_{\alpha} \ \tilde{\alpha}^\dagger \tilde{\alpha} + \omega_{\beta} \ \tilde{\beta}^\dagger \tilde{\beta} + \frac{\Omega}{2} \ \tilde{\alpha} \tilde{\beta}^\dagger  +  \frac{\Omega^*}{2} \ \tilde{\alpha}^\dagger \tilde{\beta}, \label{eq:ham1} \\
 & = & \begin{pmatrix}
 \tilde{\alpha}^\dagger & \tilde{\beta}^\dagger
 \end{pmatrix} \begin{pmatrix}
 \omega_{\alpha} & \Omega^*/2 \\ \Omega/2 & \omega_{\beta}
 \end{pmatrix} \begin{pmatrix}
 \tilde{\alpha} \\ \tilde{\beta}
 \end{pmatrix}, \\
 & = & \underline{\tilde{\alpha}}^\dagger \underline{H}_{\mathrm{in}} \underline{\tilde{\alpha}}, \label{eq:ham1mat}
\end{eqnarray} 
where $\alpha$ and $\beta$ respectively denote the spin-up and -down magnons, which constitute our preferred natural basis as discussed above in section \ref{sec:overview}. $\omega_{\alpha,\beta}$ are the energies of the uncoupled modes and $\Omega$ accounts for the coherent mode coupling. In this manuscript, we set $\hbar = 1$. We further identify operators with overhead tilde and matrices/vectors with an underline. The Hamiltonian in Eq.~(\ref{eq:ham1}) can be brought to a diagonal form:
\begin{eqnarray}
\tilde{H}  & = &  \begin{pmatrix}
\tilde{\psi}^\dagger_1 & \tilde{\psi}^\dagger_2
\end{pmatrix} \begin{pmatrix}
\omega_{1} & 0 \\ 0 & \omega_{2}
\end{pmatrix} \begin{pmatrix}
\tilde{\psi}_1 \\ \tilde{\psi}_2
\end{pmatrix}, \label{eq:ham2} \\
& = & \underline{\tilde{\psi}}^\dagger \underline{H}_{\mathrm{diag}} \underline{\tilde{\psi}},
\end{eqnarray}
via a linear transformation $\underline{\tilde{\alpha}} = \underline{P} \underline{\tilde{\psi}}$ substituted in Eq.~(\ref{eq:ham1mat}) leading to:
\begin{align}\label{eq:diag1}
\underline{H}_{\mathrm{diag}} = &  \underline{P}^\dagger \underline{H}_{\mathrm{in}} \underline{P} .
\end{align}
Since $\tilde{\alpha}$, $\tilde{\beta}$, $\tilde{\psi}_1$, and $\tilde{\psi}_2$ are annihilation operators for bosonic modes, they obey the standard commutation relations which lead to the condition:
\begin{align}\label{eq:commutation}
\underline{P} \ \underline{P}^\dagger  = & \ \underline{I} \ = \underline{P}^\dagger \ \underline{P},
\end{align} 
where $\underline{I}$ is the $2 \times 2$ identity matrix. $\underline{P}$ is thus a unitary matrix with $\underline{P}^\dagger = \underline{P}^{-1}$ making Eq.~(\ref{eq:diag1}) a standard diagonalization procedure. The latter is accomplished by solving the eigenvalue problem defined by: 
\begin{align}
\underline{H}_{\mathrm{in}} \underline{\chi} = &  \lambda \underline{\chi},
\end{align} 
where the two eigenvalues $\lambda$ become $\omega_{1,2}$ of Eq.~\eqref{eq:ham2} and the corresponding eigenvectors $\underline{\chi}_{1,2}$ make up the columns of $\underline{P}$. We may thus write explicitly:
\begin{eqnarray}
\underline{P} & = & \begin{pmatrix}
\underline{\chi}_1 & \underline{\chi}_2
\end{pmatrix} \ = \ \begin{pmatrix}
c_{1\alpha} & c_{2\alpha} \\
c_{1\beta} & c_{2\beta}
\end{pmatrix}.
\end{eqnarray}
The diagonal elements of Eq.~\eqref{eq:commutation} impose the constraints $|c_{i\alpha}|^2 + |c_{i\beta}|^2 = 1$ for $i = 1,2$, which allows us to express $\underline{P}$ via the Bloch sphere representation of the eigenvectors $\underline{\chi}_{i}$:
\begin{eqnarray}
\underline{P} & = & \begin{pmatrix}
\cos \left( \frac{\theta_1}{2} \right) &  \cos \left( \frac{\theta_2}{2} \right) \\
e^{\left( i \phi_1 \right)} \sin \left( \frac{\theta_1}{2} \right) & e^{\left( i \phi_2 \right)} \sin \left( \frac{\theta_2}{2} \right)
\end{pmatrix},
\end{eqnarray}
where $\theta_{1,2}$ and $\phi_{1,2}$ represent the two eigenvectors on the Bloch sphere~\footnote{In this representation, an overall phase factor in both the eigenvectors has been disregarded.}. The off-diagonal elements of Eq.~\eqref{eq:commutation} further lead to the condition $\theta_2 = \pi - \theta_1$ and $\phi_2 = \pi + \phi_1$, which is equivalent to reversing the direction of the Bloch sphere representation of $\underline{\chi}_1$ in obtaining that of $\underline{\chi}_2$. Employing this, the transformation matrix $\underline{P}$ can be described via two antiparallel vectors starting at the origin and ending on the unit-radius Bloch sphere (Fig.~\ref{fig:BS}) leading to the following convenient parameterization:
\begin{eqnarray}
\underline{P} & = & \begin{pmatrix}
\cos \left( \frac{\theta}{2} \right) &  - e^{\left( - i \phi \right)} \sin \left( \frac{\theta}{2} \right) \\
e^{\left( i \phi \right)} \sin \left( \frac{\theta}{2} \right) & \cos \left( \frac{\theta}{2} \right)
\end{pmatrix},
\end{eqnarray}
where the subscript 1 has been dropped from $\theta$ and $\phi$. The relation between the eigenmodes and the assumed natural basis then becomes:
\begin{eqnarray}
\underline{\tilde{\psi}} & = & \underline{P}^{-1} \underline{\tilde{\alpha}} \ = \ \underline{P}^{\dagger} \underline{\tilde{\alpha}}, \\
\underline{\tilde{\psi}}^\dagger & = &   \underline{\tilde{\alpha}}^\dagger \underline{P}, \\
\begin{pmatrix}
\tilde{\psi}_1^\dagger & \tilde{\psi}_2^\dagger
\end{pmatrix} & = &   \begin{pmatrix}
\tilde{\alpha}^\dagger & \tilde{\beta}^\dagger
\end{pmatrix} \begin{pmatrix}
\cos \left( \frac{\theta}{2} \right) &  - e^{\left( - i \phi \right)} \sin \left( \frac{\theta}{2} \right) \\
e^{\left( i \phi \right)} \sin \left( \frac{\theta}{2} \right) & \cos \left( \frac{\theta}{2} \right)
\end{pmatrix}. \label{eq:eigmodes}
\end{eqnarray}
To summarize this subsection, the transformation ($\underline{P}$) describing the eigenmodes ($\underline{\tilde{\psi}}^\dagger$) in terms of the basis ($\underline{\tilde{\alpha}}^\dagger$) can be represented via two anti-parallel unit vectors on the Bloch sphere (Fig.~\ref{fig:BS}), which are also the eigenvectors of the Hamiltonian matrix $\underline{H}_{\mathrm{in}}$ [Eq.~\eqref{eq:ham1mat}]. This is already suggestive of its relation to a spin-1/2 system that we make more explicit in the next subsection.

The Bloch sphere, depicted in Fig.~\ref{fig:BS}, is in the space of creation operators. This is because the transformation matrix $\underline{P}$, which is constituted by the two eigenvectors of the $2 \times 2$ matrix $\underline{H}_{\mathrm{in}}$ [Eq.~\eqref{eq:ham1mat}], relates the creation operators of the eigenmodes with those of the basis [Eq.~\eqref{eq:eigmodes}]. This is in contrast with the typical use of the Bloch sphere, which is to depict the wavefunctions (and not any operators) of a two-level system. Furthermore, it is customary to employ a related device - Poincar\'e sphere - in representing polarization states of classical light fields~\cite{Jones2016}. Our use of the unit sphere in representing AFI modes shares similarities and differences with both of these devices. Nevertheless, we chose to employ the Bloch sphere terminology in our representation of the AFI excitations, keeping in mind that it is not a Bloch sphere in the strict sense and represents excitation creation operators.

\begin{figure}[tb]
	\begin{center}
		\includegraphics[width=150mm]{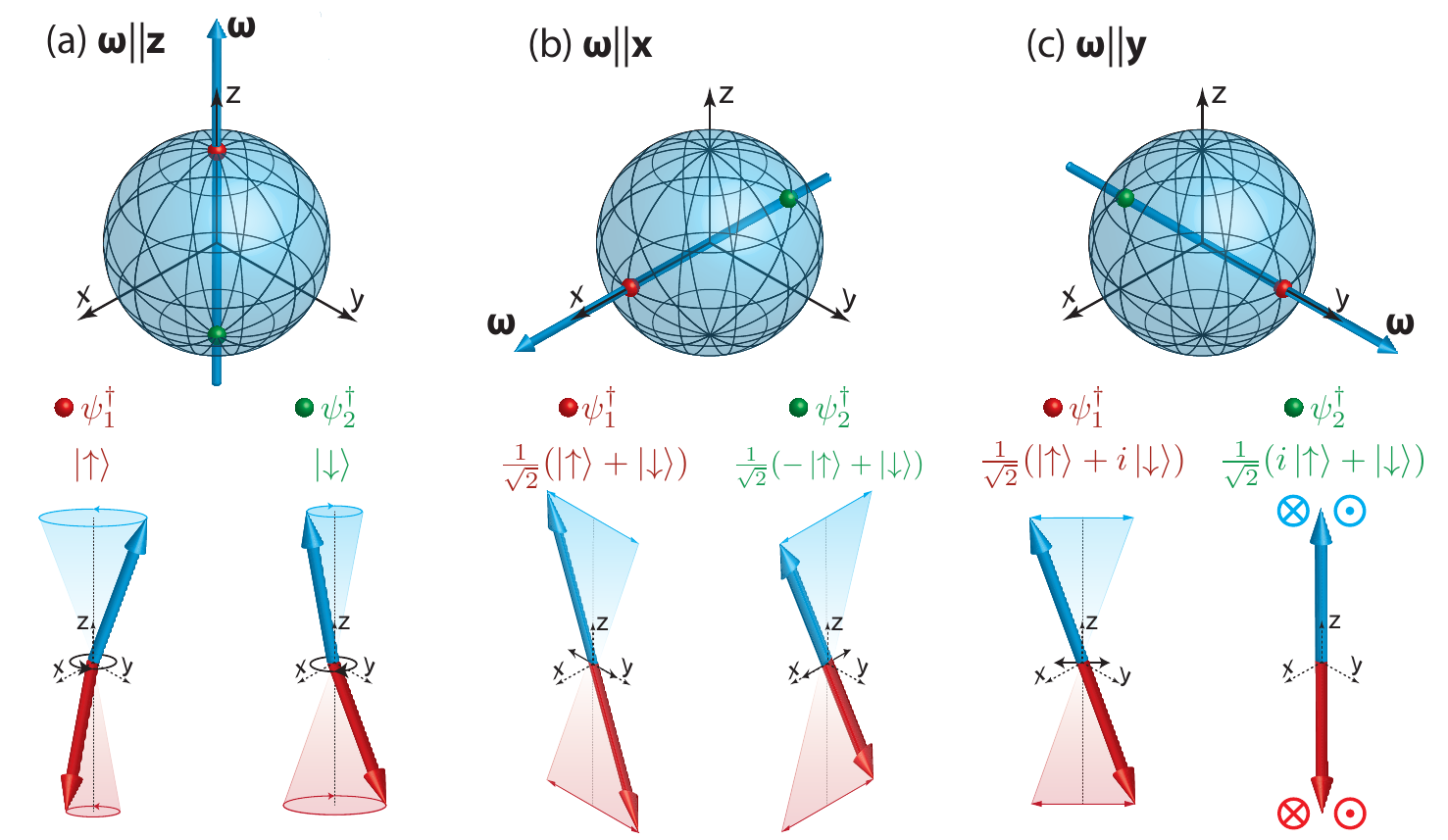}
		\caption{Schematic depiction of AFI eigenmodes for pseudofield $\pmb{\omega}$ directed along (a) $\hat{\pmb{z}}$, (b) $\hat{\pmb{x}}$, and (c) $\hat{\pmb{y}}$. In the quantum picture, the eigenmodes are expressed as superpositions of the natural Bloch sphere basis - the spin-up and -down magnons. In the Landau-Lifshitz description, the eigenmodes correspond to (a) circular precession or (b), (c) linear oscillations of the two sublattice magnetizations.}
		\label{fig:pseudoconfigs}
	\end{center}
\end{figure}

\subsection{Pseudospin, Pseudofield, and magnon spin}
Motivated by the suggestive connection between two coupled modes and spin-1/2 systems discussed above, we define the {\em pseudospin} operator $\tilde{\pmb{L}} = \tilde{L}_x \hat{\pmb{x}} + \tilde{L}_y \hat{\pmb{y}} + \tilde{L}_z \hat{\pmb{z}}$:
\begin{eqnarray}
\tilde{L}_x & = & \frac{1}{2} \left( \underline{\tilde{\alpha}}^\dagger \underline{\sigma}_{x} \underline{\tilde{\alpha}}  \right) \ = \ \frac{1}{2} \left( \tilde{\alpha} \tilde{\beta}^\dagger +  \tilde{\alpha}^\dagger \tilde{\beta} \right), \label{eq:lx} \\
\tilde{L}_y & = & \frac{1}{2} \left( \underline{\tilde{\alpha}}^\dagger \underline{\sigma}_{y} \underline{\tilde{\alpha}}  \right) \ = \ \frac{i}{2} \left( \tilde{\alpha} \tilde{\beta}^\dagger -  \tilde{\alpha}^\dagger \tilde{\beta} \right), \\
\tilde{L}_z & = & \frac{1}{2} \left( \underline{\tilde{\alpha}}^\dagger \underline{\sigma}_{z} \underline{\tilde{\alpha}}  \right) \ = \ \frac{1}{2} \left( \tilde{\alpha}^\dagger \tilde{\alpha} -  \tilde{\beta}^\dagger \tilde{\beta} \right),
\end{eqnarray}
where $\underline{\sigma}_{x,y,z}$ are the Pauli matrices. The pseudospin operator components can be shown to satisfy the standard angular momentum commutation relations: $[\tilde{L}_j , \tilde{L}_k] = i \epsilon_{jkl} \tilde{L}_{l}$. It is also convenient to define:
\begin{eqnarray}
\tilde{L}_0 & = & \frac{1}{2} \left( \underline{\tilde{\alpha}}^\dagger \underline{\sigma}_{0} \underline{\tilde{\alpha}}  \right) \ = \ \frac{1}{2} \left( \tilde{\alpha}^\dagger \tilde{\alpha} +  \tilde{\beta}^\dagger \tilde{\beta} \right), \label{eq:l}
\end{eqnarray} 
where $\underline{\sigma}_0$ is the $2 \times 2$ identity matrix. Employing Eqs.~\eqref{eq:lx}-\eqref{eq:l}, the Hamiltonian in Eq.~\eqref{eq:ham1} may be written in terms of the pseudospin operator as:
\begin{eqnarray}\label{eq:hampseudo}
\tilde{H}  & = & 2 \omega_0 \tilde{L}_0 -  \pmb{\omega} \cdot \tilde{\pmb{L}},
\end{eqnarray}
where $\omega_0$ and the components of $\pmb{\omega}$, assumed real, are given by
\begin{eqnarray}
\omega_0 & = & \frac{\omega_\alpha + \omega_\beta}{2}, \\
\omega_{z} & = & - \left( \omega_\alpha - \omega_\beta \right), \label{eq:omegaz}\\
\omega_{x} + i \omega_{y} & = & - \Omega . \label{eq:omegaxy}
\end{eqnarray}
Comparing Eq.~\eqref{eq:hampseudo} with the typical spin-1/2 Hamiltonian and employing the analysis of previous subsection, we can directly see that the Bloch vectors that characterize the eigenmodes are collinear with $\pmb{\omega}$, as depicted in Fig.~\ref{fig:BS}. The corresponding eigenmode energies become $\omega_0 \mp |\pmb{\omega}|/2$. The quantity $\pmb{\omega}$ is thus termed {\em pseudofield} as it couples to the pseudospin in a manner similar to how a magnetic field couples to an actual spin. 

Finally, we can relate the actual magnonic spin to pseudospin by recognizing that $\alpha$ and $\beta$ modes correspond to spin $+1$ and $-1$ magnons. Therefore the excitation spin operator is defined as
\begin{eqnarray}
\tilde{S} & = & \tilde{\alpha}^\dagger \tilde{\alpha} - \tilde{\beta}^\dagger \tilde{\beta}, \\
  & = &  2 \tilde{L}_z. \label{eq:magspin}
\end{eqnarray}

Knowledge of the pseudofield $\pmb{\omega}$ thus allows a simple and direct understanding of the eigenmodes in terms of the associated Bloch vectors that are collinear with the pseudofield (Fig.~\ref{fig:BS}). For $\pmb{\omega} \parallel \hat{\pmb{z}}$, the eigenmodes are same as our natural basis of spin-up and -down magnons [Fig.~\ref{fig:pseudoconfigs}(a)]. When $\pmb{\omega} \parallel \hat{\pmb{x}}$, the eigenmodes are spin-zero excitations comprising equal superpositions of $\alpha$ and $\beta$ modes [Fig.~\ref{fig:pseudoconfigs}(b)]. In Landau-Lifshitz description, the two eigenmodes correspond to linear oscillations of the N\'eel vector in two orthogonal planes. For $\pmb{\omega} \parallel \hat{\pmb{y}}$, the eigenmodes are still spin-zero excitations with different phase factors in the superposition [Fig.~\ref{fig:pseudoconfigs}(c)]. From Landau-Lifshitz dynamics perspective, the eigenmodes still correspond to linear oscillations of the N\'eel vector in two orthogonal planes, which are rotated by 45 degrees with respect to the eigenmodes corresponding to $\pmb{\omega} \parallel \hat{\pmb{x}}$.

\section{Pseudospin chemical potential}\label{sec:chemical}
In the previous section, we have defined the pseudospin operator along with other quantities and operators that allow us to describe the eigenmodes. However, we did not discuss observables and how they can be evaluated. We take up this task in the present section.

We begin by recognizing the problem at hand and comparing it to the case of electrons. The spin carried by an electronic state can be evaluated once the eigenstate is known since each state can be occupied only once - the electron is either there or not. In contrast, the bosonic modes under consideration here can bear any integer occupation numbers and thus a knowledge of the eigenmodes does not suffice in determining the physical quantities such as spin. Our considered natural basis of $\alpha$ and $\beta$ modes is spanned by the basis wavefunctions $\ket{N_{\alpha},N_{\beta}}$, where $N_{\alpha}$ and $N_{\beta}$ denote the integer number of corresponding excitations and run from 0 to $\infty$. This basis is, in principle, complete and can be used to describe any state, including those which are the eigenstates of $\psi_1$ and $\psi_2$ modes. This requires keeping track of coherent superpositions and off-diagonal elements in the density matrix describing the system. Such a representation precludes a pragmatic description in terms of quasi-equilibrium distributions and quantities, such as chemical potential, which only allow diagonal elements of the density matrix to be non-zero. 

To alleviate this problem, we employ an overcomplete basis by including the eigenstates of $\psi_1$ and $\psi_2$ for all $\theta$ and $\phi$ [Eq.~\eqref{eq:eigmodes}]. We thus define a pseudospin chemical potential vector that captures the necessary off-diagonal coherences in the density matrix via its direction and vectorial nature. The solution exploited is again motivated by the corresponding analysis of electrons~\cite{Fabian2007,Wu2010}. Simply put, our defined pseudospin chemical potential vector contains information about the eigenmodes as well as their nonequilibrium occupancy.

Let us consider eigenmodes characterized by $\theta$, $\phi$ [Eq.~\eqref{eq:eigmodes}] and assume an occupancy of $N_1$ and $N_2$. The corresponding wavefunction $\keteig{N_1,N_2}$ can, in principle, be expressed as a sum over our natural basis states $\keteigin{n_1,n_2}$, the latter being a complete basis. However, as discussed above, we employ an overcomplete basis via eigenstates corresponding to general $\theta$ and $\phi$. The pseudospin expectation value for the wavefunction $\keteig{N_1,N_2}$ is evaluated as:
\begin{eqnarray}
\langle \tilde{L}_x \rangle & = & \frac{1}{2} \left\langle \underline{\tilde{\alpha}}^\dagger \underline{\sigma}_{x} \underline{\tilde{\alpha}}   \right\rangle , \label{eq:lxexp} \\
  & = &  \frac{1}{2} \left\langle \underline{\tilde{\psi}}^\dagger \underline{P}^\dagger \underline{\sigma}_{x} \underline{P} \underline{\tilde{\psi}}   \right\rangle, \\
  & = & \frac{1}{2} \sin \theta \cos \phi (N_1 - N_2), \\
 \langle \tilde{L}_y \rangle & = & \frac{1}{2} \sin \theta \sin \phi (N_1 - N_2), \\
  \langle \tilde{L}_z \rangle & = & \frac{1}{2} \cos \theta (N_1 - N_2).
\end{eqnarray}
Motivated by our present goal of describing the nonequilibrium state via quasi-equilibrium quantities, we further assume that the two modes have the same temperature but different chemical potentials $\mu_{1,2}$ such that
\begin{eqnarray}\label{eq:n12}
N_{1,2} & = &  \left( \exp \left( \frac{\omega_{1,2} - \mu_{1,2}}{k_B T} \right) - 1  \right)^{-1} ,
\end{eqnarray}
where $k_{B}$ is the Boltzmann constant and $T$ is the temperature. Employing Eqs.~\eqref{eq:lxexp}-\eqref{eq:n12} along with $\omega_{1,2} = \omega_0 \mp |\pmb{\omega}|/2$ and $|\pmb{\omega}|,|\mu_1 - \mu_2| \ll \omega_0 $, we obtain the following expression for the pseudospin expectation value $\pmb{L} \equiv \langle \tilde{\pmb{L}} \rangle$:
\begin{eqnarray}\label{eq:pseudoexpfin}
\pmb{L} & = & \frac{1}{2} \left( \left. - \frac{\partial N(\epsilon)}{\partial \epsilon} \right|_{\epsilon = \omega_0} \right) \left( \pmb{\omega} + \pmb{\mu}_s \right),
\end{eqnarray}
where $N(\epsilon) \equiv 1/ (\exp (\epsilon/(k_B T)) - 1)$ is the Bose distribution function and we have defined the pseudospin chemical potential:
\begin{eqnarray}\label{eq:pseudochem}
\pmb{\mu}_s & \equiv &  (\mu_1 - \mu_2) \left( \sin \theta \cos \phi \hat{\pmb{x}} + \sin \theta \sin \phi \hat{\pmb{y}} + \cos \theta \hat{\pmb{z}} \right).
\end{eqnarray}
Thus, we see from Eq.~\eqref{eq:pseudoexpfin} that the pseudospin value has two contributions. The first is an equilibrium effect stemming from the energy, and thus occupancy, difference between the two eigenmodes. The second is caused by an imbalance of quasi-chemical potentials making it a nonequilibrium effect. Within the linear response considered here, both of these contributions are adequately captured by the vectors - pseudofield and pseudospin chemical potential defined via Eq.~\ref{eq:pseudochem}. The eigenmode information ($\theta,~\phi$) has conveniently been absorbed by the directions of pseudofield and pseudospin chemical potential allowing for a general description employing the natural basis. Since the magnon spin operator is proportional to $\tilde{L}_z$ [Eq.~\eqref{eq:magspin}], the typical magnon spin accumulation corresponds to the $z$ component of our pseudospin chemical potential [Eq.~\eqref{eq:pseudochem}].

\section{Dynamics}\label{sec:dynamics}
Thus far, we have largely considered the equilibrium description of the two coupled modes. Even in discussing the pseudospin chemical potential, which represents a nonequilibrium quantity, we assumed it to point along the pseudofield that determines the equilibrium modes. We now consider the situation when pseudospin is not necessarily collinear with the pseudofield.

Once again, we begin by recognizing the problem at hand and outlining the solution. Our general goal is to establish a time dependence of the pseudospin expectation value. The typical approach would be to determine the time evolution of the initial wavefunction and evaluating expectation value of the pseudospin operator using the time-dependent wavefunction. This approach is complicated by the bosonic system under consideration that allows for a large range of initial wavefunctions with different occupancies of the basis states. A convenient solution is found by working with the Heisenberg picture in which the operators themselves evolve while wavefunction and density matrix remain constant. The system dynamics can thus be captured via the operator evolution and is applicable for any initial density matrix.

The Heisenberg equation of motion for the pseudospin operator becomes:
\begin{eqnarray}
\frac{d \tilde{\pmb{L}}}{dt} & = & \frac{1}{i}  \left[ \tilde{\pmb{L}} ,  \tilde{H} \right],
\end{eqnarray}
where we employ the Hamiltonian as expressed in Eq.~\eqref{eq:hampseudo}. Using the pseudospin commutation relations: $[\tilde{L}_j , \tilde{L}_k] = i \epsilon_{jkl} \tilde{L}_{l}$, the time evolution becomes:
\begin{eqnarray}\label{eq:pseudodyn}
\frac{d \tilde{\pmb{L}}}{dt} & = & \tilde{\pmb{L}} \times \pmb{\omega},
\end{eqnarray} 
where $\pmb{\omega}$ is the pseudofield. Employing Eq.~\ref{eq:pseudoexpfin}, we obtain the dynamical equation for pseudospin chemical potential:
\begin{eqnarray}
\frac{d \pmb{\mu}_s}{dt} & = & \pmb{\mu}_s \times \pmb{\omega}.
\end{eqnarray}

\begin{figure}[tb]
	\begin{center}
		\includegraphics[width=150mm]{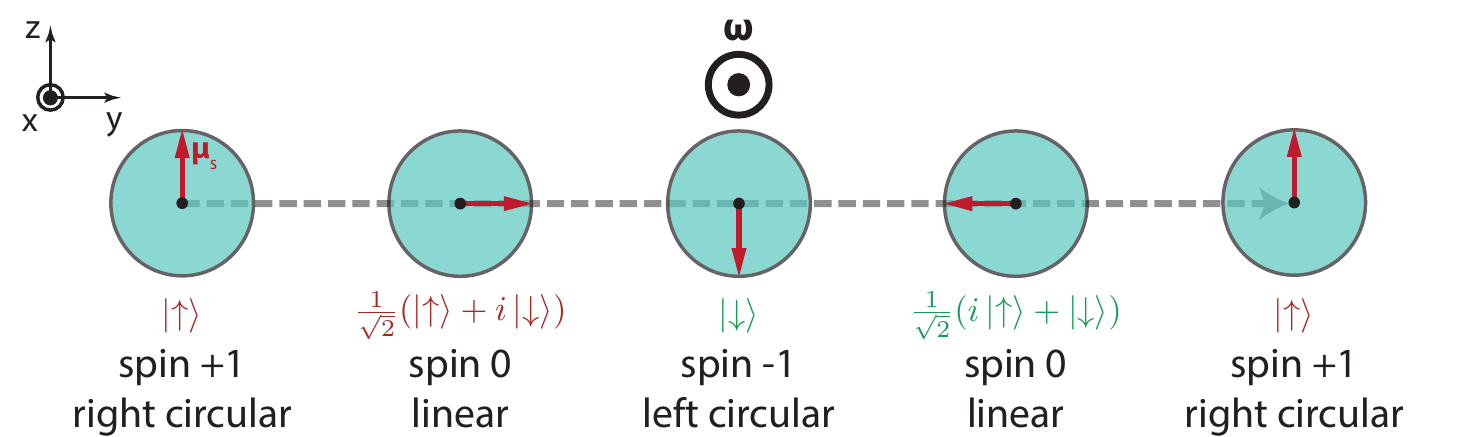}
		\caption{Schematic depicting the precession of pseudospin chemical potential vector about a pseudofield directed along $\hat{\pmb{x}}$. The analogous transmutation between the antiferromagnetic magnonic modes is also shown. The arrow of time goes from left to right.}
		\label{fig:precession}
	\end{center}
\end{figure}

We thus see that the pseudospin chemical potential precesses about the pseudofield similar to the precession of electron spin chemical potential about an applied magnetic field~\cite{Kikkawa1999,Jedema2002,Fabian2007}. A difference, however, in the sense of precession arises due to the negative gyromagnetic ratio of an electron. To gain further physical insight, we consider the situation where $\pmb{\mu}_s$ initially points along $\hat{\pmb{z}}$ while the pseudofield is directed along $\hat{\pmb{x}}$ (see Fig.~\ref{fig:precession}). This can be accomplished, for example, via a nonequilibrium injection of spin-up modes in an AFI with a hard $x$-axis anisotropy as discussed in section \ref{sec:examples} and appendix \ref{sec:MH}. The chemical potential therefore precesses about the $x$-axis in the $y$-$z$ plane resulting in a nonequilibrium transformation of the modes. In terms of the classical Landau-Lifshitz description, the polarization of the injected modes transmutes from right circular to linear to left circular to linear and back to right circular. While the modes become linearly polarized in this transmutation process, the planes of their polarization make an angle of 45 degrees with respect to the planes corresponding to the eigenmodes' polarization. The latter correspond to chemical potential pointing along the pseudofield, $\hat{\pmb{x}}$ in this case. Similar to the case of electronic spin, dephasing and decoherence relax the pseudospin chemical potential to zero while attempting to align it with the pseudofield.  

\section{Diffusive transport}\label{sec:diffusion}

In the previous sections, we have considered two coupled bosonic modes in order to define and understand the key features of pseudospin and pseudofield concepts. In an AFI, the spin-up and -down magnon modes at each wavevector constitute a pair of coupled bosonic modes. Thus, a pseudospin and pseudofield can be associated with each wavevector $\pmb{k}$. The physical response and properties of the AFI, however, may bear contribution from all wavevectors. In the present section, we consider the diffusive transport of the AFI magnon modes described in terms of pseudospin. We first resume our discussion of two coupled modes with the goal of achieving physical insights into the diffusion process. This is then generalized to include all wavevectors. We thus obtain a semi-phenomenological description of the spin and pseudospin transport in an AFI. The description thus developed explicitly assumes that mode coupling $\Omega$ is much smaller than the uncoupled mode frequencies $\omega_{\alpha,\beta}$. Since exchange interaction typically sets the dominant energy scale, this assumption is valid for AFI modes in all but a small part of the phase space near $\pmb{k} = \pmb{0}$, where it may break down. 

\subsection{Two coupled modes}
We now derive a diffusive transport equation for the pseudospin carried by two coupled modes. We employ a random walk model treating the quasiparticles represented by the modes to scatter after an average time $\tau$ while moving with a speed $v$. A detailed derivation in the context of electronic spin transport has been produced elsewhere~\cite{Fabian2007}. Following a similar procedure and exploiting Eq.~\eqref{eq:pseudodyn}, we may directly write down the pseudospin diffusion equation in three dimensions:
\begin{eqnarray}\label{eq:pseudodiff1}
\frac{\partial \pmb{L}}{\partial t} & = & D \nabla^2 \pmb{L} - \frac{\pmb{L} - \pmb{L}_0}{\tau_s} + \pmb{L} \times \pmb{\omega},
\end{eqnarray}
where $\pmb{L}$ is now the pseudospin density, $D = v^2 \tau /3$ is the diffusion coefficient, $\pmb{L}_0$ is the equilibrium pseudospin density, and $\tau_s$ is the phenomenological pseudospin relaxation time. In arriving at the equation above, we have transitioned from pseudospin to pseudospin per unit volume (density) by assuming that there is a unit density of states for the two coupled modes in our system. This transition appears more natural when we discuss modes characterized by different wavevectors in the next subsection. Employing Eq.~\ref{eq:pseudoexpfin}, we can express the pseudospin density as:
\begin{eqnarray}\label{eq:pseudoexp2}
\pmb{L} & = & \frac{1}{2V} \left( \left. - \frac{\partial N(\epsilon)}{\partial \epsilon} \right|_{\epsilon = \omega_0} \right) \left( \pmb{\omega} + \pmb{\mu}_s \right) \ = \ \pmb{L}_0 + \frac{1}{2V} \left( \left. - \frac{\partial N(\epsilon)}{\partial \epsilon} \right|_{\epsilon = \omega_0} \right) \pmb{\mu}_s ,
\end{eqnarray}
where $V$ is the sample volume and the equilibrium contribution $\pmb{L}_0~(\propto \pmb{\omega})$ has been separated. Substituting the expression above into Eq.~\ref{eq:pseudodiff1}, we obtain the corresponding diffusive transport equation in terms of the pseudospin chemical potential:
\begin{eqnarray}\label{eq:pseudodiff2}
\frac{\partial \pmb{\mu}_s}{\partial t} & = & D \nabla^2 \pmb{\mu}_s - \frac{\pmb{\mu}_s}{\overleftrightarrow{\tau}_s} + \pmb{\mu}_s \times \pmb{\omega},
\end{eqnarray}
where we have also allowed for an anisotropy in the pseudospin relaxation via a tensorial relaxation time $\overleftrightarrow{\tau}_s$ representing different values for the three components.

Thus, we have obtained a description of diffusive pseudospin transport within our toy model of two coupled modes. In the next subsection, this will be generalized to the realistic case of pseudospin transport in an AFI and provides insights into the approximations involved. We now discuss one such simplification that has already been employed in the above derivation. Our considered case of coupled bosonic modes differs from the case of electrons in two crucial ways. First, in the case at hand, the total number of quasiparticles is governed by the mode occupancy given by the Bose-Einstein distribution while the number of electrons is fixed by the density of states at the Fermi level. Second, for the case of electrons, charge neutrality imposes a spatially invariant electron density which allows a separation of spin and charge transport enabling the simple diffusive description~\cite{Fabian2007,Wu2010}. The corresponding condition for our case is $\mu_1 + \mu_2 = 0$ [Eq.~\eqref{eq:n12}] and has been invoked implicitly in achieving Eq.~\eqref{eq:pseudodiff2}. This condition is fulfilled in typical AFIs due to the strong exchange-mediated and spin-conserving magnon-magnon scattering processes as has been shown recently~\cite{Flebus2019,Shen2019,Troncoso2020}. Shen explicitly points out the equivalence of this condition to that of charge neutrality and screening in metals~\cite{Shen2019}. While this has been rigorously derived for easy-axis AFIs only which correspond to pseudospin aligned with $z$ axis, the more general result for arbitrary pseudospin directions is treated as an assumption in our analysis.   

\subsection{Contribution from all wavevectors}
In the present subsection, we discuss diffusive transport of AFI magnons in terms of their pseudospin density. For each value of the wavevector $\pmb{k}$, an AFI hosts two coupled modes with the natural basis of spin-up and -down magnons. Therefore, we may associate a pseudospin with each $\pmb{k}$ and employ the analysis developed above. As in the case of electrons~\cite{Fabian2007}, we characterize the entire magnon ensemble with a common pseudospin chemical potential $\pmb{\mu}_s$. Employing Eq.~\eqref{eq:pseudoexp2}, we may thus introduce a total pseudospin density $\pmb{\mathcal{S}}$ as the sum over all wavevectors:
\begin{eqnarray}
\pmb{\mathcal{S}} & \equiv & \sum_{\pmb{k}} \frac{1}{2V} \left( \left. - \frac{\partial N(\epsilon)}{\partial \epsilon} \right|_{\epsilon = \omega_{0\pmb{k}}} \right) \left( \pmb{\omega}_{\pmb{k}} + \pmb{\mu}_s \right), \\
  & = & \chi \left( \pmb{\omega} + \pmb{\mu}_s \right), \label{eq:pseudotot}
\end{eqnarray}
where $\omega_{0\pmb{k}}$ is the uncoupled modes energy thereby constituting the dispersion relation obtained by disregarding the coupling between the spin-up and -down magnons. Further, we have defined an effective susceptibility $\chi$ and average pseudofield $\pmb{\omega}$ as
\begin{eqnarray}
\chi & \equiv & \int \frac{d^3 k}{(2 \pi)^3} \frac{1}{2} \left( \left. - \frac{\partial N(\epsilon)}{\partial \epsilon} \right|_{\epsilon = \omega_{0\pmb{k}}} \right), \\
\pmb{\omega}  = \langle \pmb{\omega}_{\pmb{k}} \rangle_{\mathrm{BZ}}& \equiv & \frac{ \int \frac{d^3 k}{(2 \pi)^3} \pmb{\omega}_{\pmb{k}} \left( \left. - \frac{\partial N(\epsilon)}{\partial \epsilon} \right|_{\epsilon = \omega_{0\pmb{k}}} \right) }{ \int \frac{d^3 k}{(2 \pi)^3} \left( \left. - \frac{\partial N(\epsilon)}{\partial \epsilon} \right|_{\epsilon = \omega_{0\pmb{k}}} \right) }.
\end{eqnarray}
With these definitions, we may sum Eq.~\eqref{eq:pseudodiff1} over all modes thereby obtaining
\begin{eqnarray}\label{eq:pseudototdiff1}
\frac{\partial \pmb{\mathcal{S}}}{\partial t} & = & D \nabla^2 \pmb{\mathcal{S}} - \frac{\pmb{\mathcal{S}} - \pmb{\mathcal{S}}_0}{\tau_s} + \pmb{\mathcal{S}} \times \pmb{\omega},
\end{eqnarray}
where the quantities now include contribution from all wavevectors. We continue to assume wavevector-independent spin relaxation ($\tau_s$) and momentum scattering ($\tau$) times for simplicity. The diffusion constant $D$ is now given by its average value:
\begin{eqnarray}
D  = \langle D_{\pmb{k}} \rangle_{\mathrm{BZ}} & \equiv & \frac{ \int \frac{d^3 k}{(2 \pi)^3} \frac{1}{3} \tau \left( \pmb{\nabla}_{\pmb{k}} \omega_{0\pmb{k}} \right)^2 \left( \left. - \frac{\partial N(\epsilon)}{\partial \epsilon} \right|_{\epsilon = \omega_{0\pmb{k}}} \right) }{ \int \frac{d^3 k}{(2 \pi)^3} \left( \left. - \frac{\partial N(\epsilon)}{\partial \epsilon} \right|_{\epsilon = \omega_{0\pmb{k}}} \right) },
\end{eqnarray}
where $\left( \pmb{\nabla}_{\pmb{k}} \omega_{0\pmb{k}} \right)^2$ is the squared group velocity of the mode. Finally, substituting Eq.~\eqref{eq:pseudotot} into Eq.~\eqref{eq:pseudototdiff1} and allowing for tensorial pseudospin relaxation, we obtain the desired diffusion equation for the AFI pseudospin chemical potential:
\begin{eqnarray}\label{eq:pseudototdiff2}
\frac{\partial \pmb{\mu}_s}{\partial t} & = & D \nabla^2 \pmb{\mu}_s - \frac{\pmb{\mu}_s}{\overleftrightarrow{\tau}_s} + \pmb{\mu}_s \times \pmb{\omega}.
\end{eqnarray}
We note again that the usual magnon spin accumulation in AFIs is given by the $z$ component of the pseudospin chemical potential and is detected in typical non-local transport experiments. Furthermore, the pseudospin current density is obtained as:
\begin{eqnarray}\label{eq:pseudocurr}
\pmb{j}_s & = & - D \pmb{\nabla} \pmb{\mathcal{S}} \ = \ - D \chi \pmb{\nabla} \pmb{\mu}_s,
\end{eqnarray}
where the current has two directions - one associated with its flow and the other with its pseudospin component. In the equation \eqref{eq:pseudocurr} above, $\pmb{\nabla}$ provides the direction of current flow while the pseudospin direction is associated with $\pmb{\mu}_s$.

In our analysis above, we first introduced pseudospin and related quantities by considering coupling between modes at a given $\pmb{k}$. Then, we summed over all modes in achieving the diffusive transport description for the entire magnon ensemble. As a result, the averaging of various physical quantities over the Brillouin zone, denoted by $\langle \cdot \rangle_{\mathrm{BZ}}$ employs $-\partial N (\omega_{0\pmb{k}}) / \partial \epsilon$ as the weighting function. Partly due to historical reasons, it is more common to sum over all modes and then consider the mode coupling or spin, for example in the case of electrons. For that order, the weighting function would simply become $N (\omega_{0\pmb{k}})$. The macroscopic physics observed in typical experiments is expected to be relatively insensitive to the exact weighting function. This is because the two weighting functions appear similar at not too small temperatures employed in most experiments and therefore yield comparable values for the averaged quantities. Furthermore, these effective parameters, such as average pseudofield, can be extracted directly from the experimental fits without making any assumptions regarding the weighting function.

\section{Non-local spin transport in anisotropic antiferromagnets}\label{sec:examples}
\begin{figure}[tb]
	\begin{center}
		\includegraphics[width=85mm]{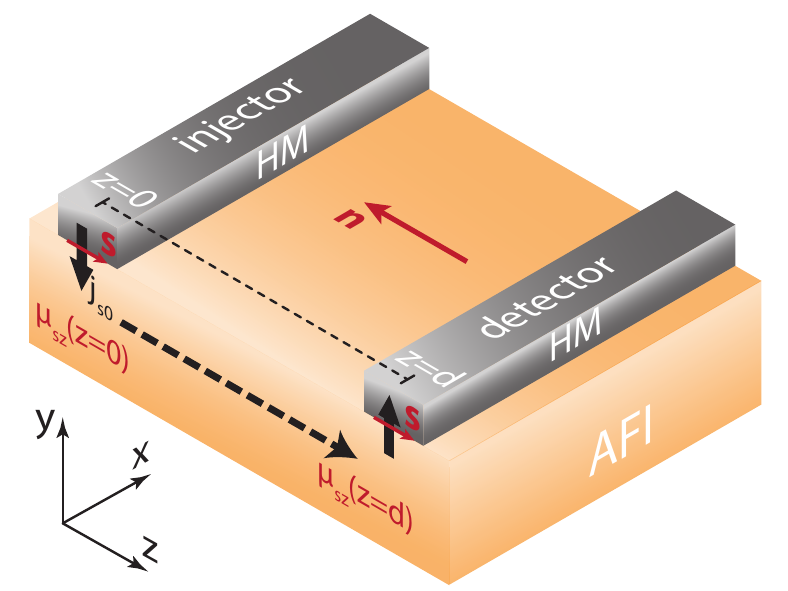}
		\caption{Device schematic for non-local magnon spin transport experiments. A $z$-polarized magnon spin and pseudospin currents are injected into and detected from the antiferromagnetic insulator (AFI) using two spatial separated heavy metal (HM) electrodes.}
		\label{fig:device}
	\end{center}
\end{figure}

A key goal of our endeavor has been a description of magnonic spin transport in N\'eel ordered AFIs with arbitrary anisotropies. This is motivated by recent experiments investigating non-local magnon transport in ferri-~\cite{CornelissenMMR,SchlitzMMR,CornelissenTemp,Shan2017,Wimmer2019} and antiferromagnets~\cite{Klaui2018,Klaui2020,Klaui_2020,Wimmer2020}. In these experiments (see Fig.~\ref{fig:device} for a schematic), a spin current is injected electrically into the magnetic insulator by the electronic spin accumulation generated in an adjacent heavy metal via the spin Hall effect~\cite{HirschSHE,Sinova2015}. The reverse mechanism allows for an electrical detection of the magnonic spin using a separate and distant detector heavy metal electrode.  In the present section, we employ the diffusive transport description developed above to investigate magnonic spin and pseudospin transport for AFIs with a varying nature of their magnonic eigenmodes. We consider a thin AFI film so that the problem is effectively one dimensional. 

\subsection{Boundary conditions}
We first introduce a simplified version of the boundary conditions relevant for the magnon spin injection and detection via heavy metal electrodes. A thorough analysis and derivation of these has been accomplished for ferromagnets~\cite{CornelissenTheory} and easy-axis AFIs~\cite{Shen2019,Troncoso2020}. In our analysis above, we have exploited the hierarchy of energy scales and treated the coherent mode coupling as a perturbation. The latter is a weaker effect, but determines the nature of AFI eigenmodes. Disregarding this coupling, AFI hosts spin-up and -down magnons as in the case of easy-axis AFIs. Thus, to the lowest (zeroth) order in mode coupling, we may carry over the boundary conditions based on easy-axis AFIs. This is tantamount to treating coupling of the AFI with electrons in the heavy metal leads to the zeroth order in the perturbation - coherent mode coupling. In other words, the coherent mode coupling has been neglected in treating the boundary conditions as it only contributes a sub-leading correction to the latter. We further work in the limit of small spin conductance of the AFI/heavy metal interfaces~\cite{CornelissenTheory,Shen2019,Troncoso2020}. Thus, the boundary conditions at the injector electrode become:
\begin{eqnarray}
- D \chi \frac{\partial \mu_{sz}}{\partial z} & = & j_{s0}, \label{eq:bc1} \\
  \frac{\partial \mu_{sx,sy}}{\partial z} & = & 0, \label{eq:bc2}
\end{eqnarray}  
where $j_{s0}$ is the magnonic spin current density injected into the AFI by the injector electrode. In typical experiments, $j_{s0}$ is proportional to the charge current driven through the injector electrode. 

The boundary conditions above imply that only spin-up and -down magnons can be injected into the AFI by a heavy metal consistent with our mathematical procedure of disregarding the mode coupling as a perturbation. A physical picture of this assumption can be painted as follows. Due to interfacial exchange interaction, an electron spin flip in the heavy metal can induce a flipping of an AFI spin localized at the interface. This delivers a localized spin of $+1$ or $-1$ to the AFI which becomes delocalized respectively into a spin-up or -down magnon mode at a time scale inversely proportional to the exchange energy. Only at a much longer time scale, inversely proportional to the mode coupling frequency, these delocalized spin-up or -down magnons recognize that they are not eigenmodes leading to pseudospin precession as captured by our diffusion equation \eqref{eq:pseudototdiff2}. Finally, we treat the detector electrode to be weakly coupled to the AFI such that it does not significantly influence the magnon transport. The inverse spin Hall effect voltage in the detector electrode is thus proportional to $\mu_{sz}$ thereby providing a measure of the magnon spin.

\subsection{One-dimensional pseudospin diffusion}
With the goal of understanding non-local magnon spin transport in AFIs hosting different kinds of eigenmodes, we aim to solve Eq.~\eqref{eq:pseudototdiff2} in one dimension and steady state:
\begin{eqnarray}\label{eq:eqsolve1}
0 & = & D \frac{\partial^2 \pmb{\mu}_s}{\partial z^2} - \frac{\pmb{\mu}_s}{\tau_s} + \pmb{\mu}_s \times \pmb{\omega},
\end{eqnarray}
where we consider $\pmb{\omega} = \omega_x \hat{\pmb{x}} + \omega_z \hat{\pmb{z}}$. We have assumed an isotropic pseudospin relaxation parameterized via time $\tau_s$. Here, as discussed in section \ref{sec:eigenmodes}, $\omega_{x} = 0$ pertains to the AFI hosting as eigenmodes spin-1 magnons corresponding to circular precession of the N\'eel vector in the Landau-Lifshitz description. On the other hand, $\omega_{z} = 0$ and $\omega_{x} \neq 0$ pertains to the AFI bearing as eigenmodes spin-zero magnons, which correspond to a linear oscillation of the N\'eel vector. In the general case of $\omega_{x,z} \neq 0$, the eigenmodes have a spin magnitude between 0 and 1 corresponding to an elliptical precession of the N\'eel vector~\cite{Kamra2017,Rezende2019}.  

\begin{figure}[tb]
	\begin{center}
		\subfloat[]{\includegraphics[width=80mm]{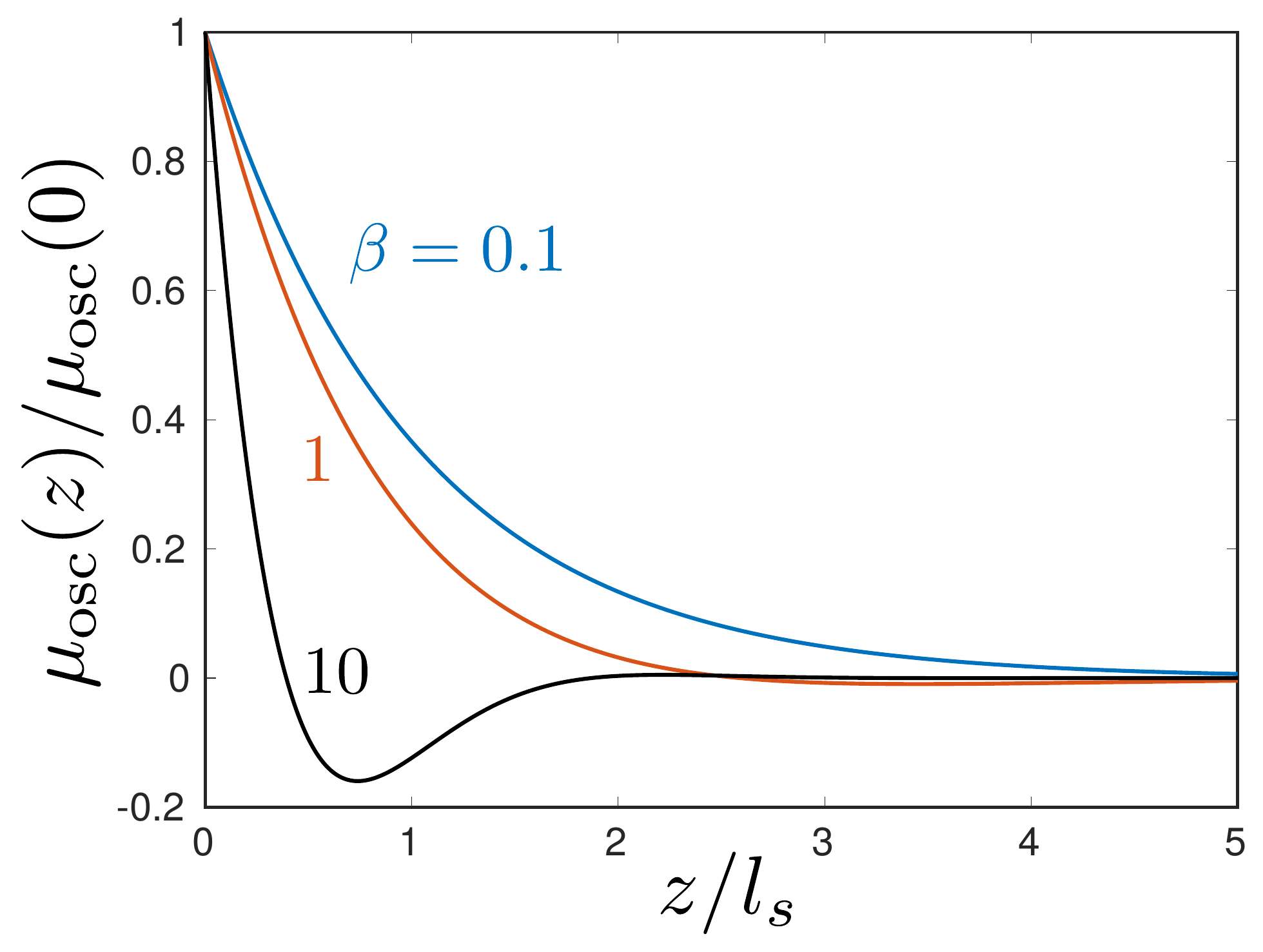}} \quad
		\subfloat[]{\includegraphics[width=80mm]{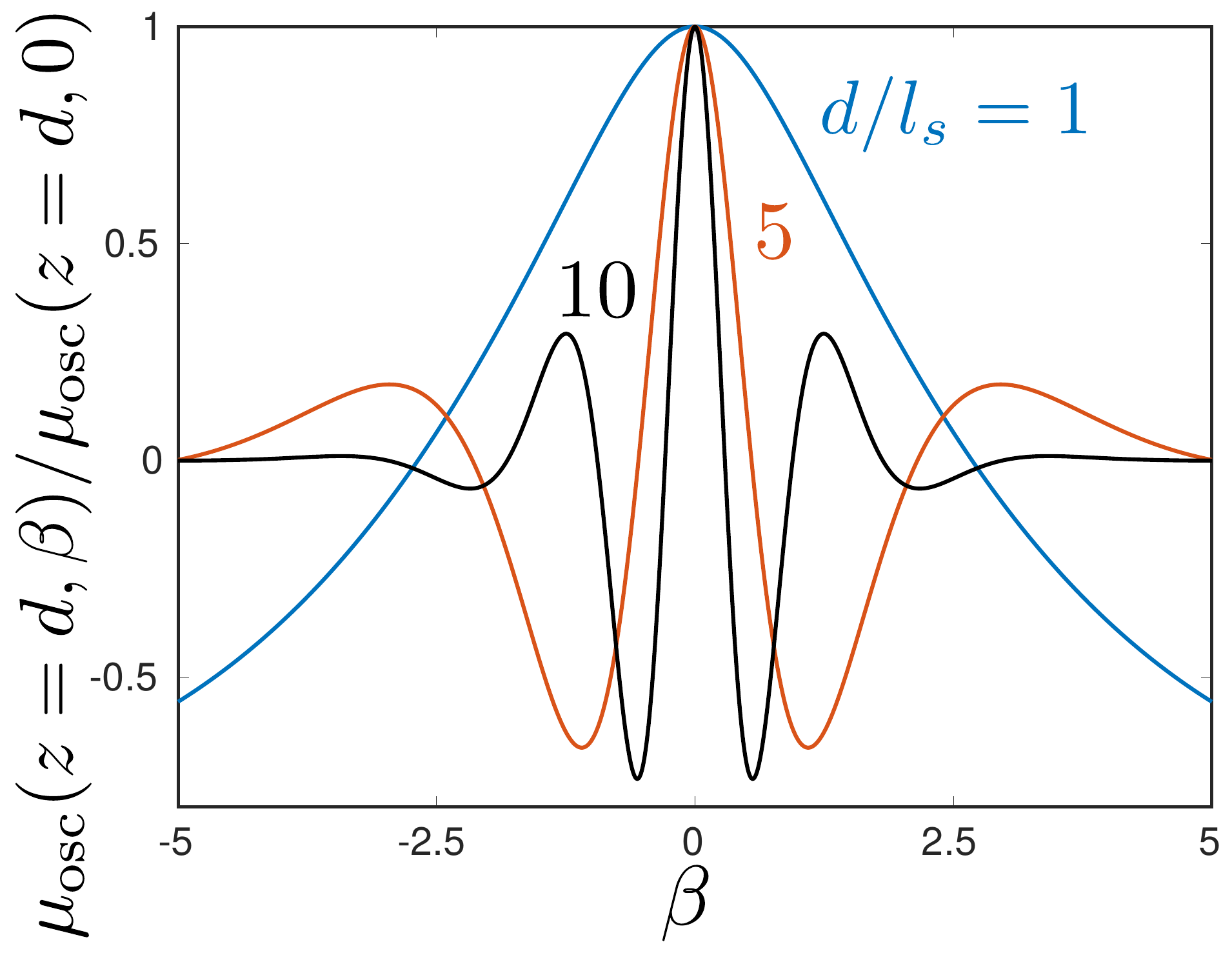}}
		\caption{Oscillating magnon spin chemical potential [Eq.~\eqref{eq:osc}] {\it vs.} (a) distance from the injector electrode and (b) normalized pseudofield magnitude. The $z$ component of the pseudospin chemical potential vector corresponds to the magnonic spin accumulation detected in a typical non-local magnon spin transport setup. In (a), we plot the appropriately normalized oscillating contribution to it $\mu_{\mathrm{osc}}$ [Eq.~\eqref{eq:osc}] as a function of the distance $z$, normalized by the spin diffusion length $l_s$, from the injector electrode for various values of the normalized pseudofield magnitude $\beta$ [Eq.~\eqref{eq:solend}]. A sign reversal in the spin accumulation, resulting from pseudospin precession about the pseudofield, occurs for $\beta \gtrsim 1$ and can be seen in the curve corresponding to $\beta = 10$. In (b), we plot a normalized magnon spin accumulation $\mu_{\mathrm{osc}}$ [Eq.~\eqref{eq:osc}] detected at an electrode a distance $d$ from the injector as a function of the pseudofield magnitude. Several oscillations can be seen for distances $d$ significantly larger than the spin diffusion length. The absolute value of $\mu_{\mathrm{osc}}$, however, diminishes with $d$ making it harder to detect multiple oscillations in experiments.}
		\label{fig:muosc}
	\end{center}
\end{figure}

We assume the injector electrode to be located at $z = 0$ and extend uniformly along the x-direction. The AFI is assumed to be thin along the y direction (Fig.~\ref{fig:device}). The problem at hand is thus one-dimensional with physical quantities varying only with $z$. Employing boundary conditions as specified by Eqs.~\eqref{eq:bc1} and \eqref{eq:bc2} along with the requirement $\pmb{\mu}_{s}(z \to \infty) \to 0$, Eq.~\eqref{eq:eqsolve1} yields the following solution after some algebra:
\begin{eqnarray}
\mu_{sz} (z) & = & \mu_{\mathrm{osc}}(z) + \mu_{\mathrm{dec}}(z),  \label{eq:solbeg}\\
 \mu_{\mathrm{osc}}(z) & = & \frac{\omega_x^2}{\omega_x^2 + \omega_z^2} \ \frac{l_s j_{s0}}{D \chi \left(a^2 + b^2\right)} \ e^{- \frac{a z}{l_s}} \left[ -b \sin \left( \frac{b z}{l_s} \right) + a \cos \left( \frac{b z}{l_s} \right) \right], \label{eq:osc}\\
 \mu_{\mathrm{dec}}(z) & = & \frac{\omega_z^2}{\omega_x^2 + \omega_z^2} \ \frac{l_s j_{s0}}{D \chi}  \ e^{- \frac{z}{l_s}}, \label{eq:dec}
\end{eqnarray}
where $l_s \equiv \sqrt{D \tau_s}$ is the spin diffusion length and we have additionally defined
\begin{align}
a \equiv & \frac{1}{\sqrt{2}} \sqrt{1 + \sqrt{1 + \beta^2}}, \label{eq:a} \\
b \equiv & \frac{1}{\sqrt{2}} \sqrt{- 1 + \sqrt{1 + \beta^2}}, \\
\beta^2  \equiv &  \tau_s^2 \left(\omega_{x}^2 + \omega_z^2  \right). \label{eq:solend}
\end{align}
Hence, we see that $\mu_{sz}$, and thus the non-local magnon spin transport signal, bears a contribution [Eq.~\eqref{eq:osc}] that oscillates with $z$ [see Fig.~\ref{fig:muosc} (a)] on account of the pseudospin precession while decaying with a characteristic length of $l_s/a$ in addition to a decaying contribution [Eq.~\eqref{eq:dec}] with the usual relaxation length of $l_s$. The solution provided by Eqs.~\eqref{eq:solbeg}-\eqref{eq:solend} allows the desired general understanding of non-local magnon transport in AFIs and constitutes a key result of this work.

\subsection{Discussion of key features}
We pause to discuss the physical content of the solution [Eqs.~\eqref{eq:solbeg}-\eqref{eq:solend}] obtained above. For $\omega_{x} = 0$, the AFI hosts spin-1 magnons that carry a diffusive spin current [Eq.~\eqref{eq:dec}] decaying with the length scale of $l_s$ defined above. This is consistent with the literature on easy-axis AFIs~\cite{Klaui2018,Shen2019,Troncoso2020}. 

For $\omega_z = 0$ and $\omega_{x} \neq 0$, spin-1 magnons are injected at $z = 0$ but they are no longer the eigenmodes, the latter being spin-zero magnons. The pseudospin therefore precesses about the pseudofield giving rise to an oscillation in the magnon spin and chemical potential arriving at the detector as described by Eq.~\eqref{eq:osc} [see Fig.~\ref{fig:muosc} (a)]~\cite{Wimmer2020}. In addition, the relaxation length is decreased by the factor $a$ [Eq.~\eqref{eq:a}] on account of destructive interference in the pseudospin from different $\pmb{k}$ modes arriving at the detector. For $\omega_z \neq 0$ and $\omega_{x} \neq 0$, the eigenmodes bear spin between 0 and 1 corresponding to elliptical precession of the N\'eel vector and the non-local signal is provided by an interplay between both oscillating [Eq.~\eqref{eq:osc}] and decaying [Eq.~\eqref{eq:dec}] contributions~\cite{Wimmer2020}. In this case, the pseudospin precession frequency is determined by the total magnitude of the pseudofield bearing contributions from $\omega_{x}$ and $\omega_z$. On the other hand, the decaying contribution to the non-local signal [Eq.~\eqref{eq:dec}] is determined essentially by the fractional circular polarization content of the eigenmodes [$\omega_z^2/ (\omega_z^2 + \omega_x^2)$].

The pseudospin precession and non-local spin transport discussed here is analogous to the case of spin precession with electrons~\cite{Kikkawa1999,Jedema2002,Fabian2007}, as anticipated at the outset. Thus, our analysis above allows an understanding of the AFI magnon pseudospin dynamics and Hanle effect observed recently~\cite{Wimmer2020}. The net pseudofield $\pmb{\omega}$ can be controlled by changing the equilibrium configuration of the AFI, e.g. via an externally applied magnetic field. Thus, the non-local signal observed at a fixed detector electrode oscillates [see Fig.~\ref{fig:muosc} (b)] as a function of an applied field and in accordance with Eqs.~\eqref{eq:solbeg}-\eqref{eq:solend}. The exact dependence of the pseudofield on an applied field depends on the microscopic details of the AFI and can be evaluated using the results obtained in our section \ref{sec:diffusion}. The observed AFI magnon Hanle effect~\cite{Wimmer2020} invokes a control of non-collinearity between the two sublattice magnetizations via an external field and thus, strictly speaking, goes beyond our analysis here restricted to collinear ground states. However, a successful accounting of the observed signal using the formalism introduced here justifies its use a posteriori. A rigorous accounting of the non-collinearity effects will be addressed elsewhere. We also note that conventions employed in the present manuscript differ significantly at various places from those in Ref.~\onlinecite{Wimmer2020}.

In obtaining the solution Eqs.~\eqref{eq:solbeg}-\eqref{eq:solend}, we have assumed isotropic pseudospin relaxation parameterized by a single time $\tau_s$ for simplicity. However, since our preferred natural basis is spin-up and -down magnons corresponding to pseudospin pointing along $z$ direction, we may expect the corresponding relaxation time $\tau_{sz}$ to be different from $\tau_{sx,sy}$ for other pseudospin components. More specifically, we may expect $\tau_{sz} > \tau_{sx,sy}$ since spin-up and -down magnons are our natural basis while other eigenmodes are formed from their coherent superpositions that are expected to suffer from additional dephasing mechanisms. Further, in the case considered here, $\tau_{sx}$ can be expected to differ from $\tau_{sy}$ as the pseudofield has a component along the $x$ direction thereby breaking the $x$-$y$ symmetry.

Solving the diffusion equation \eqref{eq:eqsolve1} above with anisotropic spin relaxation appears to be, for practical purposes, analytically intractable. However, we may use the solution Eqs.~\eqref{eq:solbeg}-\eqref{eq:solend} for the isotropic case to develop qualitative insights for the anisotropic case. Let us consider $\tau_{sx} \approx \tau_{sy} < \tau_{sz}$ which may be expected on physical grounds as discussed above. This implies that pseudospin relaxes faster when it points away from the $z$ axis. Since the decaying contribution to the non-local signal [Eq.~\eqref{eq:dec}] is mediated simply by the finite magnon spin and does not involve any pseudospin precession, it is practically unaffected by smaller dephasing times $\tau_{sx,sy}$. Therefore, this contribution continues to propagate with a length scale $l_s$. On the other hand, the oscillating contribution to the non-local signal [Eq.~\eqref{eq:osc}] is a direct consequence of pseudospin precession requiring it to deviate away from the $z$-axis, it is expected to decay even faster than $l_s/a$ and might be damped before any oscillation can be manifested.

\section{Summary}\label{sec:summary}
Taking inspiration from the coupled-boson representation of spin, we have developed a quantum field theoretic pseudospin description of the magnonic excitations in an antiferromagnetic insulator (AFI). Employing the simple case of two coherently coupled bosonic modes, we have introduced the concepts of pseudospin and pseudofield. These have been shown to provide a general and intuitive understanding of eigenmodes in an AFI. The nature of an antiferromagnetic eigenmode - circular precession or linear oscillation of N\'eel vector - has been shown to be associated with points on a Bloch sphere. The $z$ coordinate of this point pertains to the actual spin carried by the corresponding magnon mode. We have shown that in nonequilibrium situations, the pseudospin precesses about the pseudofield similar to Larmor precession of an electron spin about an applied magnetic field. This pseudospin precession corresponds to a transmutation of the antiferromagnetic modes. Employing these ideas, we have obtained a description for diffusive transport of magnonic spin in AFIs. Solving the equation thus obtained, we have delineated the qualitative features of recent experimental reports on non-local magnon spin transport in AFIs. The role of eigenmode ellipticity or spin in non-local experiments as well as the recent observation of antiferromagnetic magnon Hanle effect have been clarified. The methodology developed herein is expected to find applications in understanding magnonic spin transport in a broad range of AFIs. Due to its validity for any coherently coupled bosonic modes, it may also trigger development of spin-dynamics-inspired physical insights for, among others, coupled optomechanical~\cite{Aspelmeyer2014} and optomagnonic~\cite{MacNeill2019,Liensberger2019,Kusminskiy2016,Harder2018,Huebl2013} systems.

\section*{Acknowledgments}
We thank Siddhartha Omar, Rudolf Gross, Matthias Opel, Stephan Gepr\"ags, and Janine G\"uckelhorn for valuable discussions. We acknowledge financial support from the Research Council of Norway through its Centers of Excellence funding scheme, project 262633, ``QuSpin'' and the Deutsche Forschungsgemeinschaft (DFG, German Research Foundation) under Germany’s Excellence Strategy EXC-2111-390814868 and Project AL2110/2-1.

\appendix

\section{Magnon Hamiltonian: perturbative treatment and pseudofield}\label{sec:MH}
Considering a concrete antiferromagnetic insulator (AFI) Hamiltonian, we now elucidate the perturbative evaluation of the corresponding pseudofield. Furthermore, we wish to consider a model that allows a continuous transition between spin-1 and spin-zero magnons as the eigenmodes. Thus, we assume a two-sublattice AFI described by the Hamiltonian:
\begin{eqnarray}\label{eq:AFIham}
\tilde{H}_{\mathrm{AFI}} & = & \tilde{H}_{\mathrm{Z}} + \tilde{H}_{\mathrm{ex}} + \tilde{H}_{\mathrm{ea}} + \tilde{H}_{\mathrm{ha}},
\end{eqnarray}  
accounting for contributions from the Zeeman (Z), exchange (ex), easy-axis (ea), and hard-axis (ha) anisotropies given by:
\begin{align}
\tilde{H}_{\mathrm{Z}} & =  \mu_0 |\gamma| H_{0} \sum \left[ \tilde{S}_{Az}(\pmb{r}_i) + \tilde{S}_{Bz}(\pmb{r}_j)\right], \\
\tilde{H}_{\mathrm{ex}} & = J \sum_{\pmb{r}_i,\pmb{\delta}} \tilde{\pmb{S}}_{A}(\pmb{r}_i) \cdot \tilde{\pmb{S}}_B(\pmb{r}_i + \pmb{\delta}), \\
\tilde{H}_{\mathrm{ea}} & = - K_{\mathrm{ea}} \sum \left[ \left(\tilde{S}_{Az}(\pmb{r}_i) \right)^2 + \left(\tilde{S}_{Bz}(\pmb{r}_j) \right)^2 \right] , \\
\tilde{H}_{\mathrm{ha}} & = K_{\mathrm{ha}} \sum \left[ \left(\tilde{S}_{Ax}(\pmb{r}_i) \right)^2 + \left(\tilde{S}_{Bx}(\pmb{r}_j) \right)^2 \right] , 
\end{align}
where we continue to set $\hbar = 1$. Here, $\tilde{\pmb{S}}_{A}(\pmb{r}_i)$ [$\tilde{\pmb{S}}_{B}(\pmb{r}_j)$] is operator for the spin located at $\pmb{r}_i$ ($\pmb{r}_j$) on sublattice A (B), while $\pmb{\delta}$ denotes the vector to a nearest neighbor. We have assumed an applied magnetic field $H_0 \hat{\pmb{z}}$ and $\gamma~(<0)$ is the gyromagnetic ratio, same for both sublattices. $J~(>0)$ parameterizes the antiferromagnetic exchange interaction between the two sublattices. The positive parameters $K_{\mathrm{ea}}$ and $K_{\mathrm{ha}}$ account for anisotropies with easy-axis and hard-axis along $z$ and $x$ directions, respectively. As we will see later, the hard-axis anisotropy in this model breaks the axial symmetry and spin conservation about the $z$-axis thereby coherently coupling the spin-up and -down magnons~\cite{Kamra2017,Liensberger2019,Kamra2019}. 

We assume a N\'eel ordered ground state with the sublattice A spin pointing along $-\hat{\pmb{z}}$ while the spin for sublattice B is oriented along $\hat{\pmb{z}}$. With this assumed ground state, the linearized Holstein-Primakoff transformation~\cite{Holstein1940,Akhiezer1968} become:
\begin{align}
\tilde{S}_{A+}(\bm{r}_i) = \sqrt{2S} ~\tilde{a}_i^\dagger, \quad \tilde{S}_{A-}(\bm{r}_i) =  \sqrt{2S} ~ \tilde{a}_i, \quad \tilde{S}_{Az}(\bm{r}_i) =  - S + \tilde{a}^\dagger_i \tilde{a}_i, \label{eq:hp1} \\
\tilde{S}_{B+}(\bm{r}_j) =  \sqrt{2S}~ \tilde{b}_j, \quad \tilde{S}_{B-}(\bm{r}_j) =  \sqrt{2S} ~\tilde{b}_j^\dagger, \quad \tilde{S}_{Bz}(\bm{r}_j) =   S - \tilde{b}^\dagger_j \tilde{b}_j, \label{eq:hp2}
\end{align}
where $\tilde{S}_{A\pm} = \tilde{S}_{Ax} \pm i \tilde{S}_{Ay}$, $\tilde{S}_{B\pm} = \tilde{S}_{Bx} \pm i \tilde{S}_{By}$ and $S$ is the spin magnitude at each site. $\tilde{a}_i$ and $\tilde{b}_j$ are the magnon annihilation operators on sublattices A and B, respectively. Employing the Holstein-Primakoff transformation above and switching to Fourier space, the AFI Hamiltonian in Eq.~\eqref{eq:AFIham} is simplified to the following magnon Hamiltonian (disregarding a constant energy offset):
\begin{align}\label{eq:magtot}
\tilde{H}_{\mathrm{mag}} = & \sum_{\pmb{k}} \left[ A_{\pmb{k}} \tilde{a}^\dagger_{\pmb{k}} \tilde{a}_{\pmb{k}} + B_{\pmb{k}} \tilde{b}^\dagger_{\pmb{k}} \tilde{b}_{\pmb{k}} + \left( C_{\pmb{k}} \tilde{a}_{\pmb{k}} \tilde{b}_{-\pmb{k}} + \mathrm{h.c.} \right) +  \left( D_{\pmb{k}} \tilde{a}_{\pmb{k}} \tilde{a}_{-\pmb{k}} + E_{\pmb{k}} \tilde{b}_{\pmb{k}} \tilde{b}_{-\pmb{k}} + \mathrm{h.c.} \right)  \right],
\end{align}
where 
\begin{align}
A_{\pmb{k}} & = J S Z + 2 K_{\mathrm{ea}} S + K_{\mathrm{ha}} S + \mu_0 |\gamma|  H_0, \label{eq:ak} \\
B_{\pmb{k}} & = J S Z + 2 K_{\mathrm{ea}} S + K_{\mathrm{ha}} S - \mu_0 |\gamma|  H_0, \\
C_{\pmb{k}} & = J S Z \gamma_{\pmb{k}}, \\
D_{\pmb{k}} & = E_{\pmb{k}}  = \frac{K_{\mathrm{ha}} S}{2}, \label{eq:dk}
\end{align}  
where $Z$ is the coordination number of the lattice, and $\gamma_{\pmb{k}} \equiv  (1/Z) \sum_{\pmb{\delta}} e^{i \pmb{k} \cdot \pmb{\delta}} $ with $\pmb{\delta}$ running over nearest neighbors. 

In order to perform a perturbative analysis~\cite{Kamra2019,Liensberger2019}, we now split the total magnon Hamiltonian [Eq.~\eqref{eq:magtot}] into base and perturbation contributions $\tilde{H}_{\mathrm{mag}} = \tilde{H}_{\mathrm{base}} + \tilde{H}_{\mathrm{pert}}$ with:
\begin{eqnarray}
\tilde{H}_{\mathrm{base}} & = &  \sum_{\pmb{k}} \left[ \bar{A}_{\pmb{k}} \left( \tilde{a}^\dagger_{\pmb{k}} \tilde{a}_{\pmb{k}} +  \tilde{b}^\dagger_{\pmb{k}} \tilde{b}_{\pmb{k}} \right) + \left( C_{\pmb{k}} \tilde{a}_{\pmb{k}} \tilde{b}_{-\pmb{k}} + \mathrm{h.c.} \right) \right], \\
 \tilde{H}_{\mathrm{pert}} & = & \sum_{\pmb{k}} \left[ \Delta A_{\pmb{k}} \left( \tilde{a}^\dagger_{\pmb{k}} \tilde{a}_{\pmb{k}} - \tilde{b}^\dagger_{\pmb{k}} \tilde{b}_{\pmb{k}} \right) +  \left( D_{\pmb{k}} \tilde{a}_{\pmb{k}} \tilde{a}_{-\pmb{k}} + E_{\pmb{k}} \tilde{b}_{\pmb{k}} \tilde{b}_{-\pmb{k}} + \mathrm{h.c.} \right)\right], \label{eq:pertham1}
\end{eqnarray}
where we define $\bar{A}_{\pmb{k}} \equiv (A_{\pmb{k}} + B_{\pmb{k}})/2$ and $\Delta A_{\pmb{k}} \equiv (A_{\pmb{k}} - B_{\pmb{k}})/2$. The base Hamiltonian can be diagonalized using a Bogoliubov transformation resulting in~\cite{Kamra2019}
\begin{align}\label{eq:base}
\tilde{H}_{\mathrm{base}} & =  \sum_{\pmb{k}} \omega_{0\pmb{k}} \left( \tilde{\alpha}^\dagger_{\pmb{k}} \tilde{\alpha}_{\pmb{k}} +  \tilde{\beta}^\dagger_{\pmb{k}} \tilde{\beta}_{\pmb{k}}  \right),
\end{align}
where $\omega_{0\pmb{k}} = \sqrt{\bar{A}_{\pmb{k}}^2 - C_{\pmb{k}}^2}$ becomes the dispersion of the uncoupled modes. Here, $\alpha$ and $\beta$ modes are the spin-up and -down magnons bearing a spin along $\hat{\pmb{z}}$ of $+1$ and $-1$, respectively. These constitute our natural basis as discussed in the main text. The Bogoliubov transformation that allowed us to obtain the diagonal base Hamiltonian Eq.~\eqref{eq:base} is given by:~\cite{Kamra2019}
\begin{align}
\tilde{a}_{\pmb{k}} = & ~u_{\pmb{k}} \tilde{\alpha}_{\pmb{k}} - v_{\pmb{k}} \tilde{\beta}_{-\pmb{k}}^\dagger, \quad \tilde{b}_{\pmb{k}} = u_{\pmb{k}} \tilde{\beta}_{\pmb{k}} - v_{\pmb{k}} \tilde{\alpha}_{-\pmb{k}}^\dagger, \label{eq:magop} \\
u_{\pmb{k}} = & ~\sqrt{\frac{\bar{A}_{\pmb{k}} + \omega_{0\pmb{k}}}{2  \omega_{0\pmb{k}}}},  \ \qquad v_{\pmb{k}} = ~\sqrt{\frac{\bar{A}_{\pmb{k}} - \omega_{0\pmb{k}}}{2 \omega_{0\pmb{k}}}}. \label{eq:uv}
\end{align} 
Employing Eq.~\eqref{eq:magop} above, we may express the perturbation Hamiltonian [Eq.~\eqref{eq:pertham1}] as:
\begin{eqnarray}\label{eq:pertham2}
\tilde{H}_{\mathrm{pert}} & = & \sum_{\pmb{k}} \Delta A_{\pmb{k}} \left( \tilde{\alpha}^\dagger_{\pmb{k}} \tilde{\alpha}_{\pmb{k}} -  \tilde{\beta}^\dagger_{\pmb{k}} \tilde{\beta}_{\pmb{k}}  \right) - 4 D_{\pmb{k}} u_{\pmb{k}} v_{\pmb{k}} \left( \tilde{\alpha}_{\pmb{k}} \tilde{\beta}_{\pmb{k}}^\dagger + \tilde{\alpha}_{\pmb{k}}^\dagger \tilde{\beta}_{\pmb{k}} \right).
\end{eqnarray}
In obtaining Eq.~\eqref{eq:pertham2} above, we have exploited the relation $D_{\pmb{k}} = E_{\pmb{k}}$ [Eq.~\eqref{eq:dk}] and the inversion symmetry of the problem, i.e. all coefficients such as $A_{\pmb{k}}, D_{\pmb{k}}, \cdots$ remain the same on replacing $\pmb{k}$ with $- \pmb{k}$. Furthermore, we have disregarded the terms, such as $\sim \tilde{\alpha}_{\pmb{k}} \tilde{\alpha}_{-\pmb{k}}$, that do not conserve the excitation number thereby making the rotating wave approximation. Employing Eqs.~\eqref{eq:omegaz} and \eqref{eq:omegaxy}, the pseudofield $\pmb{\omega}_{\pmb{k}}$ can be read off from Eq.~\eqref{eq:pertham2} as:
\begin{eqnarray}
\pmb{\omega}_{\pmb{k}} & = & 8 D_{\pmb{k}} u_{\pmb{k}} v_{\pmb{k}} \ \hat{\pmb{x}} - 2 \Delta A_{\pmb{k}} \  \hat{\pmb{z}}, \\
 & = & 4 K_{\mathrm{ha}} S u_{\pmb{k}} v_{\pmb{k}}  \ \hat{\pmb{x}} - 2 \mu_0 |\gamma| H_0 \  \hat{\pmb{z}}, \label{eq:pseudofieldex}
\end{eqnarray}
where we have employed Eqs.~\eqref{eq:ak}-\eqref{eq:dk} in simplifying the expression above. 

Thus we see that the pseudofield [Eq.~\eqref{eq:pseudofieldex}] bears a $\pmb{k}$-dependent $x$ component resulting from the hard-axis anisotropy that breaks the axial symmetry about the N\'eel vector ($z$-axis). This contribution depends on the base Hamiltonian via the factors $u_{\pmb{k}}, ~v_{\pmb{k}}$ [Eq.~\ref{eq:uv}] and decreases with an increasing wavenumber. As a result, its contribution $\omega_x$ to the pseudofield averaged over all modes is expected to decrease with an increasing temperature, since the thermally excited modes have larger wavevectors on an average at higher temperatures. This is reminiscent of a similar argument presented by Han and coworkers in explaining the temperature dependence of the spin diffusion length observed in their experiments~\cite{Han2020}. The pseudofield [Eq.~\eqref{eq:pseudofieldex}] also bears a $\pmb{k}$-independent contribution parallel to the $z$-axis which stems from the applied field and therefore, can be controlled directly. Thus, for the model AFI considered in this section [Eq.~\eqref{eq:AFIham}], it appears easy to tune the relative strengths of the pseudofield components and thus, the nature of eigenmodes hosted. Together with the results discussed in section \ref{sec:examples} [Eqs.~\eqref{eq:solbeg}-\eqref{eq:solend}], this could enable an experimental investigation of non-local magnon spin transport with continuously varying nature of the eigenmodes.

\bibliography{pseudospin.bib}

\begin{thebibliography}{65}%
\makeatletter
\providecommand \@ifxundefined [1]{%
 \@ifx{#1\undefined}
}%
\providecommand \@ifnum [1]{%
 \ifnum #1\expandafter \@firstoftwo
 \else \expandafter \@secondoftwo
 \fi
}%
\providecommand \@ifx [1]{%
 \ifx #1\expandafter \@firstoftwo
 \else \expandafter \@secondoftwo
 \fi
}%
\providecommand \natexlab [1]{#1}%
\providecommand \enquote  [1]{``#1''}%
\providecommand \bibnamefont  [1]{#1}%
\providecommand \bibfnamefont [1]{#1}%
\providecommand \citenamefont [1]{#1}%
\providecommand \href@noop [0]{\@secondoftwo}%
\providecommand \href [0]{\begingroup \@sanitize@url \@href}%
\providecommand \@href[1]{\@@startlink{#1}\@@href}%
\providecommand \@@href[1]{\endgroup#1\@@endlink}%
\providecommand \@sanitize@url [0]{\catcode `\\12\catcode `\$12\catcode
  `\&12\catcode `\#12\catcode `\^12\catcode `\_12\catcode `\%12\relax}%
\providecommand \@@startlink[1]{}%
\providecommand \@@endlink[0]{}%
\providecommand \url  [0]{\begingroup\@sanitize@url \@url }%
\providecommand \@url [1]{\endgroup\@href {#1}{\urlprefix }}%
\providecommand \urlprefix  [0]{URL }%
\providecommand \Eprint [0]{\href }%
\providecommand \doibase [0]{http://dx.doi.org/}%
\providecommand \selectlanguage [0]{\@gobble}%
\providecommand \bibinfo  [0]{\@secondoftwo}%
\providecommand \bibfield  [0]{\@secondoftwo}%
\providecommand \translation [1]{[#1]}%
\providecommand \BibitemOpen [0]{}%
\providecommand \bibitemStop [0]{}%
\providecommand \bibitemNoStop [0]{.\EOS\space}%
\providecommand \EOS [0]{\spacefactor3000\relax}%
\providecommand \BibitemShut  [1]{\csname bibitem#1\endcsname}%
\let\auto@bib@innerbib\@empty
\bibitem [{\citenamefont {Bauer}\ \emph {et~al.}(2012)\citenamefont {Bauer},
  \citenamefont {Saitoh},\ and\ \citenamefont {van Wees}}]{Bauer2012}%
  \BibitemOpen
  \bibfield  {author} {\bibinfo {author} {\bibfnamefont {Gerrit E.~W.}\
  \bibnamefont {Bauer}}, \bibinfo {author} {\bibfnamefont {Eiji}\ \bibnamefont
  {Saitoh}}, \ and\ \bibinfo {author} {\bibfnamefont {Bart~J.}\ \bibnamefont
  {van Wees}},\ }\bibfield  {title} {\enquote {\bibinfo {title} {Spin
  caloritronics},}\ }\href {\doibase 10.1038/nmat3301} {\bibfield  {journal}
  {\bibinfo  {journal} {Nature Materials}\ }\textbf {\bibinfo {volume} {11}},\
  \bibinfo {pages} {391--399} (\bibinfo {year} {2012})}\BibitemShut {NoStop}%
\bibitem [{\citenamefont {Chumak}\ \emph {et~al.}(2015)\citenamefont {Chumak},
  \citenamefont {Vasyuchka}, \citenamefont {Serga},\ and\ \citenamefont
  {Hillebrands}}]{Chumak2015}%
  \BibitemOpen
  \bibfield  {author} {\bibinfo {author} {\bibfnamefont {A.~V.}\ \bibnamefont
  {Chumak}}, \bibinfo {author} {\bibfnamefont {V.~I.}\ \bibnamefont
  {Vasyuchka}}, \bibinfo {author} {\bibfnamefont {A.~A.}\ \bibnamefont
  {Serga}}, \ and\ \bibinfo {author} {\bibfnamefont {B.}~\bibnamefont
  {Hillebrands}},\ }\bibfield  {title} {\enquote {\bibinfo {title} {Magnon
  spintronics},}\ }\href {\doibase 10.1038/nphys3347} {\bibfield  {journal}
  {\bibinfo  {journal} {Nature Physics}\ }\textbf {\bibinfo {volume} {11}},\
  \bibinfo {pages} {453--461} (\bibinfo {year} {2015})}\BibitemShut {NoStop}%
\bibitem [{\citenamefont {Kajiwara}\ \emph {et~al.}(2010)\citenamefont
  {Kajiwara}, \citenamefont {Harii}, \citenamefont {Takahashi}, \citenamefont
  {Ohe}, \citenamefont {Uchida}, \citenamefont {Mizuguchi}, \citenamefont
  {Umezawa}, \citenamefont {Kawai}, \citenamefont {Ando}, \citenamefont
  {Takanashi}, \citenamefont {Maekawa},\ and\ \citenamefont
  {Saitoh}}]{KajiwaraNL}%
  \BibitemOpen
  \bibfield  {author} {\bibinfo {author} {\bibfnamefont {Y.}~\bibnamefont
  {Kajiwara}}, \bibinfo {author} {\bibfnamefont {K.}~\bibnamefont {Harii}},
  \bibinfo {author} {\bibfnamefont {S.}~\bibnamefont {Takahashi}}, \bibinfo
  {author} {\bibfnamefont {J.}~\bibnamefont {Ohe}}, \bibinfo {author}
  {\bibfnamefont {K.}~\bibnamefont {Uchida}}, \bibinfo {author} {\bibfnamefont
  {M.}~\bibnamefont {Mizuguchi}}, \bibinfo {author} {\bibfnamefont
  {H.}~\bibnamefont {Umezawa}}, \bibinfo {author} {\bibfnamefont
  {H.}~\bibnamefont {Kawai}}, \bibinfo {author} {\bibfnamefont
  {K.}~\bibnamefont {Ando}}, \bibinfo {author} {\bibfnamefont {K.}~\bibnamefont
  {Takanashi}}, \bibinfo {author} {\bibfnamefont {S.}~\bibnamefont {Maekawa}},
  \ and\ \bibinfo {author} {\bibfnamefont {E.}~\bibnamefont {Saitoh}},\
  }\bibfield  {title} {\enquote {\bibinfo {title} {Transmission of electrical
  signals by spin-wave interconversion in a magnetic insulator},}\ }\href
  {\doibase 10.1038/nature08876} {\bibfield  {journal} {\bibinfo  {journal}
  {Nature}\ }\textbf {\bibinfo {volume} {464}},\ \bibinfo {pages} {262--266}
  (\bibinfo {year} {2010})}\BibitemShut {NoStop}%
\bibitem [{\citenamefont {Baltz}\ \emph {et~al.}(2018)\citenamefont {Baltz},
  \citenamefont {Manchon}, \citenamefont {Tsoi}, \citenamefont {Moriyama},
  \citenamefont {Ono},\ and\ \citenamefont {Tserkovnyak}}]{Baltz2018}%
  \BibitemOpen
  \bibfield  {author} {\bibinfo {author} {\bibfnamefont {V.}~\bibnamefont
  {Baltz}}, \bibinfo {author} {\bibfnamefont {A.}~\bibnamefont {Manchon}},
  \bibinfo {author} {\bibfnamefont {M.}~\bibnamefont {Tsoi}}, \bibinfo {author}
  {\bibfnamefont {T.}~\bibnamefont {Moriyama}}, \bibinfo {author}
  {\bibfnamefont {T.}~\bibnamefont {Ono}}, \ and\ \bibinfo {author}
  {\bibfnamefont {Y.}~\bibnamefont {Tserkovnyak}},\ }\bibfield  {title}
  {\enquote {\bibinfo {title} {Antiferromagnetic spintronics},}\ }\href
  {\doibase 10.1103/revmodphys.90.015005} {\bibfield  {journal} {\bibinfo
  {journal} {Reviews of Modern Physics}\ }\textbf {\bibinfo {volume} {90}},\
  \bibinfo {pages} {015005} (\bibinfo {year} {2018})}\BibitemShut {NoStop}%
\bibitem [{\citenamefont {Chumak}\ \emph {et~al.}(2014)\citenamefont {Chumak},
  \citenamefont {Serga},\ and\ \citenamefont {Hillebrands}}]{HillebrandsMag}%
  \BibitemOpen
  \bibfield  {author} {\bibinfo {author} {\bibfnamefont {Andrii~V.}\
  \bibnamefont {Chumak}}, \bibinfo {author} {\bibfnamefont {Alexander.~A.}\
  \bibnamefont {Serga}}, \ and\ \bibinfo {author} {\bibfnamefont {Burkard}\
  \bibnamefont {Hillebrands}},\ }\bibfield  {title} {\enquote {\bibinfo {title}
  {Magnon transistor for all-magnon data processing},}\ }\href
  {http://dx.doi.org/10.1038/ncomms5700} {\bibfield  {journal} {\bibinfo
  {journal} {Nature Communications}\ }\textbf {\bibinfo {volume} {5}} (\bibinfo
  {year} {2014})}\BibitemShut {NoStop}%
\bibitem [{\citenamefont {Ganzhorn}\ \emph {et~al.}(2016)\citenamefont
  {Ganzhorn}, \citenamefont {Klingler}, \citenamefont {Wimmer}, \citenamefont
  {Gepr\"{a}gs}, \citenamefont {Gross}, \citenamefont {Huebl},\ and\
  \citenamefont {Goennenwein}}]{KathrinLogik}%
  \BibitemOpen
  \bibfield  {author} {\bibinfo {author} {\bibfnamefont {Kathrin}\ \bibnamefont
  {Ganzhorn}}, \bibinfo {author} {\bibfnamefont {Stefan}\ \bibnamefont
  {Klingler}}, \bibinfo {author} {\bibfnamefont {Tobias}\ \bibnamefont
  {Wimmer}}, \bibinfo {author} {\bibfnamefont {Stephan}\ \bibnamefont
  {Gepr\"{a}gs}}, \bibinfo {author} {\bibfnamefont {Rudolf}\ \bibnamefont
  {Gross}}, \bibinfo {author} {\bibfnamefont {Hans}\ \bibnamefont {Huebl}}, \
  and\ \bibinfo {author} {\bibfnamefont {Sebastian T.~B.}\ \bibnamefont
  {Goennenwein}},\ }\bibfield  {title} {\enquote {\bibinfo {title}
  {Magnon-based logic in a multi-terminal {YIG}/pt nanostructure},}\ }\href
  {\doibase 10.1063/1.4958893} {\bibfield  {journal} {\bibinfo  {journal}
  {Applied Physics Letters}\ }\textbf {\bibinfo {volume} {109}},\ \bibinfo
  {pages} {022405} (\bibinfo {year} {2016})}\BibitemShut {NoStop}%
\bibitem [{\citenamefont {Uchida}\ \emph {et~al.}(2010)\citenamefont {Uchida},
  \citenamefont {Xiao}, \citenamefont {Adachi}, \citenamefont {Ohe},
  \citenamefont {Takahashi}, \citenamefont {Ieda}, \citenamefont {Ota},
  \citenamefont {Kajiwara}, \citenamefont {Umezawa}, \citenamefont {Kawai},
  \citenamefont {Bauer}, \citenamefont {Maekawa},\ and\ \citenamefont
  {Saitoh}}]{UchidaSSE}%
  \BibitemOpen
  \bibfield  {author} {\bibinfo {author} {\bibfnamefont {K.}~\bibnamefont
  {Uchida}}, \bibinfo {author} {\bibfnamefont {J.}~\bibnamefont {Xiao}},
  \bibinfo {author} {\bibfnamefont {H.}~\bibnamefont {Adachi}}, \bibinfo
  {author} {\bibfnamefont {J.}~\bibnamefont {Ohe}}, \bibinfo {author}
  {\bibfnamefont {S.}~\bibnamefont {Takahashi}}, \bibinfo {author}
  {\bibfnamefont {J.}~\bibnamefont {Ieda}}, \bibinfo {author} {\bibfnamefont
  {T.}~\bibnamefont {Ota}}, \bibinfo {author} {\bibfnamefont {Y.}~\bibnamefont
  {Kajiwara}}, \bibinfo {author} {\bibfnamefont {H.}~\bibnamefont {Umezawa}},
  \bibinfo {author} {\bibfnamefont {H.}~\bibnamefont {Kawai}}, \bibinfo
  {author} {\bibfnamefont {G.~E.~W.}\ \bibnamefont {Bauer}}, \bibinfo {author}
  {\bibfnamefont {S.}~\bibnamefont {Maekawa}}, \ and\ \bibinfo {author}
  {\bibfnamefont {E.}~\bibnamefont {Saitoh}},\ }\bibfield  {title} {\enquote
  {\bibinfo {title} {Spin seebeck insulator},}\ }\href {\doibase
  10.1038/nmat2856} {\bibfield  {journal} {\bibinfo  {journal} {Nature
  Materials}\ }\textbf {\bibinfo {volume} {9}},\ \bibinfo {pages} {894--897}
  (\bibinfo {year} {2010})}\BibitemShut {NoStop}%
\bibitem [{\citenamefont {Althammer}(2018)}]{Althammer2018}%
  \BibitemOpen
  \bibfield  {author} {\bibinfo {author} {\bibfnamefont {Matthias}\
  \bibnamefont {Althammer}},\ }\bibfield  {title} {\enquote {\bibinfo {title}
  {Pure spin currents in magnetically ordered insulator/normal metal
  heterostructures},}\ }\href {\doibase 10.1088/1361-6463/aaca89} {\bibfield
  {journal} {\bibinfo  {journal} {Journal of Physics D: Applied Physics}\
  }\textbf {\bibinfo {volume} {51}},\ \bibinfo {pages} {313001} (\bibinfo
  {year} {2018})}\BibitemShut {NoStop}%
\bibitem [{\citenamefont {Mook}\ \emph {et~al.}(2014)\citenamefont {Mook},
  \citenamefont {Henk},\ and\ \citenamefont {Mertig}}]{Mook2014}%
  \BibitemOpen
  \bibfield  {author} {\bibinfo {author} {\bibfnamefont {Alexander}\
  \bibnamefont {Mook}}, \bibinfo {author} {\bibfnamefont {J\"urgen}\
  \bibnamefont {Henk}}, \ and\ \bibinfo {author} {\bibfnamefont {Ingrid}\
  \bibnamefont {Mertig}},\ }\bibfield  {title} {\enquote {\bibinfo {title}
  {Edge states in topological magnon insulators},}\ }\href {\doibase
  10.1103/PhysRevB.90.024412} {\bibfield  {journal} {\bibinfo  {journal} {Phys.
  Rev. B}\ }\textbf {\bibinfo {volume} {90}},\ \bibinfo {pages} {024412}
  (\bibinfo {year} {2014})}\BibitemShut {NoStop}%
\bibitem [{\citenamefont {Mook}\ \emph {et~al.}(2017)\citenamefont {Mook},
  \citenamefont {G\"{o}bel}, \citenamefont {Henk},\ and\ \citenamefont
  {Mertig}}]{Mook2017}%
  \BibitemOpen
  \bibfield  {author} {\bibinfo {author} {\bibfnamefont {Alexander}\
  \bibnamefont {Mook}}, \bibinfo {author} {\bibfnamefont {B\"{o}rge}\
  \bibnamefont {G\"{o}bel}}, \bibinfo {author} {\bibfnamefont {J\"{u}rgen}\
  \bibnamefont {Henk}}, \ and\ \bibinfo {author} {\bibfnamefont {Ingrid}\
  \bibnamefont {Mertig}},\ }\bibfield  {title} {\enquote {\bibinfo {title}
  {Magnon transport in noncollinear spin textures: Anisotropies and topological
  magnon hall effects},}\ }\href {\doibase 10.1103/physrevb.95.020401}
  {\bibfield  {journal} {\bibinfo  {journal} {Physical Review B}\ }\textbf
  {\bibinfo {volume} {95}},\ \bibinfo {pages} {020401} (\bibinfo {year}
  {2017})}\BibitemShut {NoStop}%
\bibitem [{\citenamefont {Nakata}\ \emph {et~al.}(2017)\citenamefont {Nakata},
  \citenamefont {Simon},\ and\ \citenamefont {Loss}}]{Nakata2017}%
  \BibitemOpen
  \bibfield  {author} {\bibinfo {author} {\bibfnamefont {Kouki}\ \bibnamefont
  {Nakata}}, \bibinfo {author} {\bibfnamefont {Pascal}\ \bibnamefont {Simon}},
  \ and\ \bibinfo {author} {\bibfnamefont {Daniel}\ \bibnamefont {Loss}},\
  }\bibfield  {title} {\enquote {\bibinfo {title} {Spin currents and magnon
  dynamics in insulating magnets},}\ }\href {\doibase 10.1088/1361-6463/aa5b09}
  {\bibfield  {journal} {\bibinfo  {journal} {Journal of Physics D: Applied
  Physics}\ }\textbf {\bibinfo {volume} {50}},\ \bibinfo {pages} {114004}
  (\bibinfo {year} {2017})}\BibitemShut {NoStop}%
\bibitem [{\citenamefont {Onose}\ \emph {et~al.}(2010)\citenamefont {Onose},
  \citenamefont {Ideue}, \citenamefont {Katsura}, \citenamefont {Shiomi},
  \citenamefont {Nagaosa},\ and\ \citenamefont {Tokura}}]{Onose2010}%
  \BibitemOpen
  \bibfield  {author} {\bibinfo {author} {\bibfnamefont {Y.}~\bibnamefont
  {Onose}}, \bibinfo {author} {\bibfnamefont {T.}~\bibnamefont {Ideue}},
  \bibinfo {author} {\bibfnamefont {H.}~\bibnamefont {Katsura}}, \bibinfo
  {author} {\bibfnamefont {Y.}~\bibnamefont {Shiomi}}, \bibinfo {author}
  {\bibfnamefont {N.}~\bibnamefont {Nagaosa}}, \ and\ \bibinfo {author}
  {\bibfnamefont {Y.}~\bibnamefont {Tokura}},\ }\bibfield  {title} {\enquote
  {\bibinfo {title} {Observation of the magnon hall effect},}\ }\href {\doibase
  10.1126/science.1188260} {\bibfield  {journal} {\bibinfo  {journal}
  {Science}\ }\textbf {\bibinfo {volume} {329}},\ \bibinfo {pages} {297--299}
  (\bibinfo {year} {2010})}\BibitemShut {NoStop}%
\bibitem [{\citenamefont {Ohnuma}\ \emph {et~al.}(2013)\citenamefont {Ohnuma},
  \citenamefont {Adachi}, \citenamefont {Saitoh},\ and\ \citenamefont
  {Maekawa}}]{OhnumaSSE}%
  \BibitemOpen
  \bibfield  {author} {\bibinfo {author} {\bibfnamefont {Yuichi}\ \bibnamefont
  {Ohnuma}}, \bibinfo {author} {\bibfnamefont {Hiroto}\ \bibnamefont {Adachi}},
  \bibinfo {author} {\bibfnamefont {Eiji}\ \bibnamefont {Saitoh}}, \ and\
  \bibinfo {author} {\bibfnamefont {Sadamichi}\ \bibnamefont {Maekawa}},\
  }\bibfield  {title} {\enquote {\bibinfo {title} {Spin seebeck effect in
  antiferromagnets and compensated ferrimagnets},}\ }\href {\doibase
  10.1103/PhysRevB.87.014423} {\bibfield  {journal} {\bibinfo  {journal}
  {Physical Review B}\ }\textbf {\bibinfo {volume} {87}},\ \bibinfo {pages}
  {014423} (\bibinfo {year} {2013})}\BibitemShut {NoStop}%
\bibitem [{\citenamefont {Flebus}\ \emph {et~al.}(2016)\citenamefont {Flebus},
  \citenamefont {Bender}, \citenamefont {Tserkovnyak},\ and\ \citenamefont
  {Duine}}]{Flebus2016}%
  \BibitemOpen
  \bibfield  {author} {\bibinfo {author} {\bibfnamefont {B.}~\bibnamefont
  {Flebus}}, \bibinfo {author} {\bibfnamefont {S.~A.}\ \bibnamefont {Bender}},
  \bibinfo {author} {\bibfnamefont {Y.}~\bibnamefont {Tserkovnyak}}, \ and\
  \bibinfo {author} {\bibfnamefont {R.~A.}\ \bibnamefont {Duine}},\ }\bibfield
  {title} {\enquote {\bibinfo {title} {Two-fluid theory for spin superfluidity
  in magnetic insulators},}\ }\href {\doibase 10.1103/PhysRevLett.116.117201}
  {\bibfield  {journal} {\bibinfo  {journal} {Phys. Rev. Lett.}\ }\textbf
  {\bibinfo {volume} {116}},\ \bibinfo {pages} {117201} (\bibinfo {year}
  {2016})}\BibitemShut {NoStop}%
\bibitem [{\citenamefont {Hirsch}(1999)}]{HirschSHE}%
  \BibitemOpen
  \bibfield  {author} {\bibinfo {author} {\bibfnamefont {J.~E.}\ \bibnamefont
  {Hirsch}},\ }\bibfield  {title} {\enquote {\bibinfo {title} {Spin hall
  effect},}\ }\href {\doibase 10.1103/PhysRevLett.83.1834} {\bibfield
  {journal} {\bibinfo  {journal} {Physical Review Letters}\ }\textbf {\bibinfo
  {volume} {83}},\ \bibinfo {pages} {1834--1837} (\bibinfo {year}
  {1999})}\BibitemShut {NoStop}%
\bibitem [{\citenamefont {Saitoh}\ \emph {et~al.}(2006)\citenamefont {Saitoh},
  \citenamefont {Ueda}, \citenamefont {Miyajima},\ and\ \citenamefont
  {Tatara}}]{SaitohISHE}%
  \BibitemOpen
  \bibfield  {author} {\bibinfo {author} {\bibfnamefont {E.}~\bibnamefont
  {Saitoh}}, \bibinfo {author} {\bibfnamefont {M.}~\bibnamefont {Ueda}},
  \bibinfo {author} {\bibfnamefont {H.}~\bibnamefont {Miyajima}}, \ and\
  \bibinfo {author} {\bibfnamefont {G.}~\bibnamefont {Tatara}},\ }\bibfield
  {title} {\enquote {\bibinfo {title} {Conversion of spin current into charge
  current at room temperature: Inverse spin-hall effect},}\ }\href {\doibase
  10.1063/1.2199473} {\bibfield  {journal} {\bibinfo  {journal} {Applied
  Physics Letters}\ }\textbf {\bibinfo {volume} {88}},\ \bibinfo {pages}
  {182509--182509--3} (\bibinfo {year} {2006})}\BibitemShut {NoStop}%
\bibitem [{\citenamefont {Sinova}\ \emph {et~al.}(2015)\citenamefont {Sinova},
  \citenamefont {Valenzuela}, \citenamefont {Wunderlich}, \citenamefont
  {Back},\ and\ \citenamefont {Jungwirth}}]{Sinova2015}%
  \BibitemOpen
  \bibfield  {author} {\bibinfo {author} {\bibfnamefont {Jairo}\ \bibnamefont
  {Sinova}}, \bibinfo {author} {\bibfnamefont {Sergio~O.}\ \bibnamefont
  {Valenzuela}}, \bibinfo {author} {\bibfnamefont {J.}~\bibnamefont
  {Wunderlich}}, \bibinfo {author} {\bibfnamefont {C.{\hspace{0.167em}}H.}\
  \bibnamefont {Back}}, \ and\ \bibinfo {author} {\bibfnamefont
  {T.}~\bibnamefont {Jungwirth}},\ }\bibfield  {title} {\enquote {\bibinfo
  {title} {Spin {Hall} effects},}\ }\href {\doibase 10.1103/revmodphys.87.1213}
  {\bibfield  {journal} {\bibinfo  {journal} {Reviews of Modern Physics}\
  }\textbf {\bibinfo {volume} {87}},\ \bibinfo {pages} {1213--1260} (\bibinfo
  {year} {2015})}\BibitemShut {NoStop}%
\bibitem [{\citenamefont {Valenzuela}\ and\ \citenamefont
  {Tinkham}(2006)}]{Valenzuela}%
  \BibitemOpen
  \bibfield  {author} {\bibinfo {author} {\bibfnamefont {Sergio~O}\
  \bibnamefont {Valenzuela}}\ and\ \bibinfo {author} {\bibfnamefont
  {M}~\bibnamefont {Tinkham}},\ }\bibfield  {title} {\enquote {\bibinfo {title}
  {Direct electronic measurement of the spin hall effect},}\ }\href
  {http://arxiv.org/abs/cond-mat/0605423} {\bibfield  {journal} {\bibinfo
  {journal} {Nature}\ }\textbf {\bibinfo {volume} {442}},\ \bibinfo {pages} {5}
  (\bibinfo {year} {2006})}\BibitemShut {NoStop}%
\bibitem [{Note1()}]{Note1}%
  \BibitemOpen
  \bibinfo {note} {Here, we use the term ``ferromagnet'' in a general sense to
  include ferrimagnets, such as yttrium iron garnet. Many of the experiments
  have been carried out on the latter material, while the theoretical models
  often treat it as a ferromagnet, for simplicity.}\BibitemShut {Stop}%
\bibitem [{\citenamefont {Cornelissen}\ \emph {et~al.}(2015)\citenamefont
  {Cornelissen}, \citenamefont {Liu}, \citenamefont {Duine}, \citenamefont
  {Youssef},\ and\ \citenamefont {van Wees}}]{CornelissenMMR}%
  \BibitemOpen
  \bibfield  {author} {\bibinfo {author} {\bibfnamefont {L.~J.}\ \bibnamefont
  {Cornelissen}}, \bibinfo {author} {\bibfnamefont {J.}~\bibnamefont {Liu}},
  \bibinfo {author} {\bibfnamefont {R.~A.}\ \bibnamefont {Duine}}, \bibinfo
  {author} {\bibfnamefont {J.~Ben}\ \bibnamefont {Youssef}}, \ and\ \bibinfo
  {author} {\bibfnamefont {B.~J.}\ \bibnamefont {van Wees}},\ }\bibfield
  {title} {\enquote {\bibinfo {title} {Long-distance transport of magnon spin
  information in a magnetic insulator at room~temperature},}\ }\href {\doibase
  10.1038/nphys3465} {\bibfield  {journal} {\bibinfo  {journal} {Nature
  Physics}\ }\textbf {\bibinfo {volume} {11}},\ \bibinfo {pages} {1022--1026}
  (\bibinfo {year} {2015})}\BibitemShut {NoStop}%
\bibitem [{\citenamefont {Goennenwein}\ \emph {et~al.}(2015)\citenamefont
  {Goennenwein}, \citenamefont {Schlitz}, \citenamefont {Pernpeintner},
  \citenamefont {Ganzhorn}, \citenamefont {Althammer}, \citenamefont {Gross},\
  and\ \citenamefont {Huebl}}]{SchlitzMMR}%
  \BibitemOpen
  \bibfield  {author} {\bibinfo {author} {\bibfnamefont {Sebastian T.~B.}\
  \bibnamefont {Goennenwein}}, \bibinfo {author} {\bibfnamefont {Richard}\
  \bibnamefont {Schlitz}}, \bibinfo {author} {\bibfnamefont {Matthias}\
  \bibnamefont {Pernpeintner}}, \bibinfo {author} {\bibfnamefont {Kathrin}\
  \bibnamefont {Ganzhorn}}, \bibinfo {author} {\bibfnamefont {Matthias}\
  \bibnamefont {Althammer}}, \bibinfo {author} {\bibfnamefont {Rudolf}\
  \bibnamefont {Gross}}, \ and\ \bibinfo {author} {\bibfnamefont {Hans}\
  \bibnamefont {Huebl}},\ }\bibfield  {title} {\enquote {\bibinfo {title}
  {Non-local magnetoresistance in yig/pt nanostructures},}\ }\href {\doibase
  http://dx.doi.org/10.1063/1.4935074} {\bibfield  {journal} {\bibinfo
  {journal} {Applied Physics Letters}\ }\textbf {\bibinfo {volume} {107}},\
  \bibinfo {eid} {172405} (\bibinfo {year} {2015})}\BibitemShut {NoStop}%
\bibitem [{\citenamefont {Wimmer}\ \emph {et~al.}(2019)\citenamefont {Wimmer},
  \citenamefont {Althammer}, \citenamefont {Liensberger}, \citenamefont
  {Vlietstra}, \citenamefont {Gepr\"ags}, \citenamefont {Weiler}, \citenamefont
  {Gross},\ and\ \citenamefont {Huebl}}]{Wimmer2019}%
  \BibitemOpen
  \bibfield  {author} {\bibinfo {author} {\bibfnamefont {T.}~\bibnamefont
  {Wimmer}}, \bibinfo {author} {\bibfnamefont {M.}~\bibnamefont {Althammer}},
  \bibinfo {author} {\bibfnamefont {L.}~\bibnamefont {Liensberger}}, \bibinfo
  {author} {\bibfnamefont {N.}~\bibnamefont {Vlietstra}}, \bibinfo {author}
  {\bibfnamefont {S.}~\bibnamefont {Gepr\"ags}}, \bibinfo {author}
  {\bibfnamefont {M.}~\bibnamefont {Weiler}}, \bibinfo {author} {\bibfnamefont
  {R.}~\bibnamefont {Gross}}, \ and\ \bibinfo {author} {\bibfnamefont
  {H.}~\bibnamefont {Huebl}},\ }\bibfield  {title} {\enquote {\bibinfo {title}
  {Spin transport in a magnetic insulator with zero effective damping},}\
  }\href {\doibase 10.1103/PhysRevLett.123.257201} {\bibfield  {journal}
  {\bibinfo  {journal} {Phys. Rev. Lett.}\ }\textbf {\bibinfo {volume} {123}},\
  \bibinfo {pages} {257201} (\bibinfo {year} {2019})}\BibitemShut {NoStop}%
\bibitem [{\citenamefont {Cornelissen}\ \emph
  {et~al.}(2016{\natexlab{a}})\citenamefont {Cornelissen}, \citenamefont
  {Peters}, \citenamefont {Bauer}, \citenamefont {Duine},\ and\ \citenamefont
  {van Wees}}]{CornelissenTheory}%
  \BibitemOpen
  \bibfield  {author} {\bibinfo {author} {\bibfnamefont {L.~J.}\ \bibnamefont
  {Cornelissen}}, \bibinfo {author} {\bibfnamefont {K.~J.~H.}\ \bibnamefont
  {Peters}}, \bibinfo {author} {\bibfnamefont {G.~E.~W.}\ \bibnamefont
  {Bauer}}, \bibinfo {author} {\bibfnamefont {R.~A.}\ \bibnamefont {Duine}}, \
  and\ \bibinfo {author} {\bibfnamefont {B.~J.}\ \bibnamefont {van Wees}},\
  }\bibfield  {title} {\enquote {\bibinfo {title} {Magnon spin transport driven
  by the magnon chemical potential in a magnetic insulator},}\ }\href {\doibase
  10.1103/PhysRevB.94.014412} {\bibfield  {journal} {\bibinfo  {journal}
  {Physical Review B}\ }\textbf {\bibinfo {volume} {94}},\ \bibinfo {pages}
  {014412} (\bibinfo {year} {2016}{\natexlab{a}})}\BibitemShut {NoStop}%
\bibitem [{\citenamefont {Zhang}\ and\ \citenamefont
  {Zhang}(2012)}]{ZhangMMR1}%
  \BibitemOpen
  \bibfield  {author} {\bibinfo {author} {\bibfnamefont {Steven S.-L.}\
  \bibnamefont {Zhang}}\ and\ \bibinfo {author} {\bibfnamefont {Shufeng}\
  \bibnamefont {Zhang}},\ }\bibfield  {title} {\enquote {\bibinfo {title}
  {Magnon mediated electric current drag across a ferromagnetic insulator
  layer},}\ }\href {\doibase 10.1103/PhysRevLett.109.096603} {\bibfield
  {journal} {\bibinfo  {journal} {Physical Review Letters}\ }\textbf {\bibinfo
  {volume} {109}},\ \bibinfo {pages} {096603} (\bibinfo {year}
  {2012})}\BibitemShut {NoStop}%
\bibitem [{\citenamefont {Li}\ \emph {et~al.}(2016)\citenamefont {Li},
  \citenamefont {Xu}, \citenamefont {Aldosary}, \citenamefont {Tang},
  \citenamefont {Lin}, \citenamefont {Zhang}, \citenamefont {Lake},\ and\
  \citenamefont {Shi}}]{Li2016}%
  \BibitemOpen
  \bibfield  {author} {\bibinfo {author} {\bibfnamefont {Junxue}\ \bibnamefont
  {Li}}, \bibinfo {author} {\bibfnamefont {Yadong}\ \bibnamefont {Xu}},
  \bibinfo {author} {\bibfnamefont {Mohammed}\ \bibnamefont {Aldosary}},
  \bibinfo {author} {\bibfnamefont {Chi}\ \bibnamefont {Tang}}, \bibinfo
  {author} {\bibfnamefont {Zhisheng}\ \bibnamefont {Lin}}, \bibinfo {author}
  {\bibfnamefont {Shufeng}\ \bibnamefont {Zhang}}, \bibinfo {author}
  {\bibfnamefont {Roger}\ \bibnamefont {Lake}}, \ and\ \bibinfo {author}
  {\bibfnamefont {Jing}\ \bibnamefont {Shi}},\ }\bibfield  {title} {\enquote
  {\bibinfo {title} {Observation of magnon-mediated current drag in pt/yttrium
  iron garnet/pt(ta) trilayers},}\ }\href {\doibase 10.1038/ncomms10858}
  {\bibfield  {journal} {\bibinfo  {journal} {Nature Communications}\ }\textbf
  {\bibinfo {volume} {7}},\ \bibinfo {pages} {10858} (\bibinfo {year}
  {2016})}\BibitemShut {NoStop}%
\bibitem [{\citenamefont {Lebrun}\ \emph {et~al.}(2018)\citenamefont {Lebrun},
  \citenamefont {Ross}, \citenamefont {Bender}, \citenamefont {Qaiumzadeh},
  \citenamefont {Baldrati}, \citenamefont {Cramer}, \citenamefont {Brataas},
  \citenamefont {Duine},\ and\ \citenamefont {Kl\"{a}ui}}]{Klaui2018}%
  \BibitemOpen
  \bibfield  {author} {\bibinfo {author} {\bibfnamefont {R.}~\bibnamefont
  {Lebrun}}, \bibinfo {author} {\bibfnamefont {A.}~\bibnamefont {Ross}},
  \bibinfo {author} {\bibfnamefont {S.~A.}\ \bibnamefont {Bender}}, \bibinfo
  {author} {\bibfnamefont {A.}~\bibnamefont {Qaiumzadeh}}, \bibinfo {author}
  {\bibfnamefont {L.}~\bibnamefont {Baldrati}}, \bibinfo {author}
  {\bibfnamefont {J.}~\bibnamefont {Cramer}}, \bibinfo {author} {\bibfnamefont
  {A.}~\bibnamefont {Brataas}}, \bibinfo {author} {\bibfnamefont {R.~A.}\
  \bibnamefont {Duine}}, \ and\ \bibinfo {author} {\bibfnamefont
  {M.}~\bibnamefont {Kl\"{a}ui}},\ }\bibfield  {title} {\enquote {\bibinfo
  {title} {Tunable long-distance spin transport in a crystalline
  antiferromagnetic iron oxide},}\ }\href {\doibase 10.1038/s41586-018-0490-7}
  {\bibfield  {journal} {\bibinfo  {journal} {Nature}\ }\textbf {\bibinfo
  {volume} {561}},\ \bibinfo {pages} {222--225} (\bibinfo {year}
  {2018})}\BibitemShut {NoStop}%
\bibitem [{\citenamefont {Ross}\ \emph
  {et~al.}(2020{\natexlab{a}})\citenamefont {Ross}, \citenamefont {Lebrun},
  \citenamefont {Gomonay}, \citenamefont {Grave}, \citenamefont {Kay},
  \citenamefont {Baldrati}, \citenamefont {Becker}, \citenamefont {Qaiumzadeh},
  \citenamefont {Ulloa}, \citenamefont {Jakob}, \citenamefont {Kronast},
  \citenamefont {Sinova}, \citenamefont {Duine}, \citenamefont {Brataas},
  \citenamefont {Rothschild},\ and\ \citenamefont {Kläui}}]{Klaui2020}%
  \BibitemOpen
  \bibfield  {author} {\bibinfo {author} {\bibfnamefont {Andrew}\ \bibnamefont
  {Ross}}, \bibinfo {author} {\bibfnamefont {Romain}\ \bibnamefont {Lebrun}},
  \bibinfo {author} {\bibfnamefont {Olena}\ \bibnamefont {Gomonay}}, \bibinfo
  {author} {\bibfnamefont {Daniel~A.}\ \bibnamefont {Grave}}, \bibinfo {author}
  {\bibfnamefont {Asaf}\ \bibnamefont {Kay}}, \bibinfo {author} {\bibfnamefont
  {Lorenzo}\ \bibnamefont {Baldrati}}, \bibinfo {author} {\bibfnamefont {Sven}\
  \bibnamefont {Becker}}, \bibinfo {author} {\bibfnamefont {Alireza}\
  \bibnamefont {Qaiumzadeh}}, \bibinfo {author} {\bibfnamefont {Camilo}\
  \bibnamefont {Ulloa}}, \bibinfo {author} {\bibfnamefont {Gerhard}\
  \bibnamefont {Jakob}}, \bibinfo {author} {\bibfnamefont {Florian}\
  \bibnamefont {Kronast}}, \bibinfo {author} {\bibfnamefont {Jairo}\
  \bibnamefont {Sinova}}, \bibinfo {author} {\bibfnamefont {Rembert}\
  \bibnamefont {Duine}}, \bibinfo {author} {\bibfnamefont {Arne}\ \bibnamefont
  {Brataas}}, \bibinfo {author} {\bibfnamefont {Avner}\ \bibnamefont
  {Rothschild}}, \ and\ \bibinfo {author} {\bibfnamefont {Mathias}\
  \bibnamefont {Kläui}},\ }\bibfield  {title} {\enquote {\bibinfo {title}
  {Propagation length of antiferromagnetic magnons governed by domain
  configurations},}\ }\href {\doibase 10.1021/acs.nanolett.9b03837} {\bibfield
  {journal} {\bibinfo  {journal} {Nano Letters}\ }\textbf {\bibinfo {volume}
  {20}},\ \bibinfo {pages} {306--313} (\bibinfo {year}
  {2020}{\natexlab{a}})}\BibitemShut {NoStop}%
\bibitem [{\citenamefont {Shen}(2019)}]{Shen2019}%
  \BibitemOpen
  \bibfield  {author} {\bibinfo {author} {\bibfnamefont {Ka}~\bibnamefont
  {Shen}},\ }\bibfield  {title} {\enquote {\bibinfo {title} {Pure spin current
  in antiferromagnetic insulators},}\ }\href {\doibase
  10.1103/physrevb.100.094423} {\bibfield  {journal} {\bibinfo  {journal}
  {Physical Review B}\ }\textbf {\bibinfo {volume} {100}},\ \bibinfo {pages}
  {094423} (\bibinfo {year} {2019})}\BibitemShut {NoStop}%
\bibitem [{\citenamefont {Troncoso}\ \emph {et~al.}(2020)\citenamefont
  {Troncoso}, \citenamefont {Bender}, \citenamefont {Brataas},\ and\
  \citenamefont {Duine}}]{Troncoso2020}%
  \BibitemOpen
  \bibfield  {author} {\bibinfo {author} {\bibfnamefont {Roberto~E.}\
  \bibnamefont {Troncoso}}, \bibinfo {author} {\bibfnamefont {Scott~A.}\
  \bibnamefont {Bender}}, \bibinfo {author} {\bibfnamefont {Arne}\ \bibnamefont
  {Brataas}}, \ and\ \bibinfo {author} {\bibfnamefont {Rembert~A.}\
  \bibnamefont {Duine}},\ }\bibfield  {title} {\enquote {\bibinfo {title} {Spin
  transport in thick insulating antiferromagnetic films},}\ }\href {\doibase
  10.1103/physrevb.101.054404} {\bibfield  {journal} {\bibinfo  {journal}
  {Physical Review B}\ }\textbf {\bibinfo {volume} {101}},\ \bibinfo {pages}
  {054404} (\bibinfo {year} {2020})}\BibitemShut {NoStop}%
\bibitem [{\citenamefont {Kamra}\ \emph {et~al.}(2017)\citenamefont {Kamra},
  \citenamefont {Agrawal},\ and\ \citenamefont {Belzig}}]{Kamra2017}%
  \BibitemOpen
  \bibfield  {author} {\bibinfo {author} {\bibfnamefont {Akashdeep}\
  \bibnamefont {Kamra}}, \bibinfo {author} {\bibfnamefont {Utkarsh}\
  \bibnamefont {Agrawal}}, \ and\ \bibinfo {author} {\bibfnamefont {Wolfgang}\
  \bibnamefont {Belzig}},\ }\bibfield  {title} {\enquote {\bibinfo {title}
  {Noninteger-spin magnonic excitations in untextured magnets},}\ }\href
  {\doibase 10.1103/PhysRevB.96.020411} {\bibfield  {journal} {\bibinfo
  {journal} {Phys. Rev. B}\ }\textbf {\bibinfo {volume} {96}},\ \bibinfo
  {pages} {020411} (\bibinfo {year} {2017})}\BibitemShut {NoStop}%
\bibitem [{\citenamefont {Rezende}\ \emph {et~al.}(2019)\citenamefont
  {Rezende}, \citenamefont {Azevedo},\ and\ \citenamefont
  {Rodr{\'{\i}}guez-Su{\'{a}}rez}}]{Rezende2019}%
  \BibitemOpen
  \bibfield  {author} {\bibinfo {author} {\bibfnamefont {Sergio~M.}\
  \bibnamefont {Rezende}}, \bibinfo {author} {\bibfnamefont {Antonio}\
  \bibnamefont {Azevedo}}, \ and\ \bibinfo {author} {\bibfnamefont
  {Roberto~L.}\ \bibnamefont {Rodr{\'{\i}}guez-Su{\'{a}}rez}},\ }\bibfield
  {title} {\enquote {\bibinfo {title} {Introduction to antiferromagnetic
  magnons},}\ }\href {\doibase 10.1063/1.5109132} {\bibfield  {journal}
  {\bibinfo  {journal} {Journal of Applied Physics}\ }\textbf {\bibinfo
  {volume} {126}},\ \bibinfo {pages} {151101} (\bibinfo {year}
  {2019})}\BibitemShut {NoStop}%
\bibitem [{\citenamefont {Lebrun}\ \emph {et~al.}(2020)\citenamefont {Lebrun},
  \citenamefont {Ross}, \citenamefont {Gomonay}, \citenamefont {Baltz},
  \citenamefont {Ebels}, \citenamefont {Barra}, \citenamefont {Qaiumzadeh},
  \citenamefont {Brataas}, \citenamefont {Sinova},\ and\ \citenamefont
  {Kl\"{a}ui}}]{Klaui_2020}%
  \BibitemOpen
  \bibfield  {author} {\bibinfo {author} {\bibfnamefont {Romain}\ \bibnamefont
  {Lebrun}}, \bibinfo {author} {\bibfnamefont {Andrew}\ \bibnamefont {Ross}},
  \bibinfo {author} {\bibfnamefont {Olena}\ \bibnamefont {Gomonay}}, \bibinfo
  {author} {\bibfnamefont {Vincent}\ \bibnamefont {Baltz}}, \bibinfo {author}
  {\bibfnamefont {Ursula}\ \bibnamefont {Ebels}}, \bibinfo {author}
  {\bibfnamefont {Anne~Laure}\ \bibnamefont {Barra}}, \bibinfo {author}
  {\bibfnamefont {Alireza}\ \bibnamefont {Qaiumzadeh}}, \bibinfo {author}
  {\bibfnamefont {Arne}\ \bibnamefont {Brataas}}, \bibinfo {author}
  {\bibfnamefont {Jairo}\ \bibnamefont {Sinova}}, \ and\ \bibinfo {author}
  {\bibfnamefont {Mathias}\ \bibnamefont {Kl\"{a}ui}},\ }\href@noop {}
  {\enquote {\bibinfo {title} {Long-distance spin-transport across the morin
  phase transition up to room temperature in the ultra-low damping alpha-fe2o3
  antiferromagnet},}\ } (\bibinfo {year} {2020}),\ \Eprint
  {http://arxiv.org/abs/arXiv:2005.14414} {arXiv:2005.14414} \BibitemShut
  {NoStop}%
\bibitem [{\citenamefont {Han}\ \emph {et~al.}(2020)\citenamefont {Han},
  \citenamefont {Zhang}, \citenamefont {Bi}, \citenamefont {Fan}, \citenamefont
  {Safi}, \citenamefont {Xiang}, \citenamefont {Finley}, \citenamefont {Fu},
  \citenamefont {Cheng},\ and\ \citenamefont {Liu}}]{Han2020}%
  \BibitemOpen
  \bibfield  {author} {\bibinfo {author} {\bibfnamefont {Jiahao}\ \bibnamefont
  {Han}}, \bibinfo {author} {\bibfnamefont {Pengxiang}\ \bibnamefont {Zhang}},
  \bibinfo {author} {\bibfnamefont {Zhen}\ \bibnamefont {Bi}}, \bibinfo
  {author} {\bibfnamefont {Yabin}\ \bibnamefont {Fan}}, \bibinfo {author}
  {\bibfnamefont {Taqiyyah~S.}\ \bibnamefont {Safi}}, \bibinfo {author}
  {\bibfnamefont {Junxiang}\ \bibnamefont {Xiang}}, \bibinfo {author}
  {\bibfnamefont {Joseph}\ \bibnamefont {Finley}}, \bibinfo {author}
  {\bibfnamefont {Liang}\ \bibnamefont {Fu}}, \bibinfo {author} {\bibfnamefont
  {Ran}\ \bibnamefont {Cheng}}, \ and\ \bibinfo {author} {\bibfnamefont
  {Luqiao}\ \bibnamefont {Liu}},\ }\bibfield  {title} {\enquote {\bibinfo
  {title} {Birefringence-like spin transport via linearly polarized
  antiferromagnetic magnons},}\ }\href {\doibase 10.1038/s41565-020-0703-8}
  {\bibfield  {journal} {\bibinfo  {journal} {Nature Nanotechnology}\ }\textbf
  {\bibinfo {volume} {15}},\ \bibinfo {pages} {563--568} (\bibinfo {year}
  {2020})}\BibitemShut {NoStop}%
\bibitem [{\citenamefont {Wimmer}\ \emph {et~al.}(2020)\citenamefont {Wimmer},
  \citenamefont {Kamra}, \citenamefont {G\"uckelhorn}, \citenamefont {Opel},
  \citenamefont {Gepr\"ags}, \citenamefont {Gross}, \citenamefont {Huebl},\
  and\ \citenamefont {Althammer}}]{Wimmer2020}%
  \BibitemOpen
  \bibfield  {author} {\bibinfo {author} {\bibfnamefont {Tobias}\ \bibnamefont
  {Wimmer}}, \bibinfo {author} {\bibfnamefont {Akashdeep}\ \bibnamefont
  {Kamra}}, \bibinfo {author} {\bibfnamefont {Janine}\ \bibnamefont
  {G\"uckelhorn}}, \bibinfo {author} {\bibfnamefont {Matthias}\ \bibnamefont
  {Opel}}, \bibinfo {author} {\bibfnamefont {Stephan}\ \bibnamefont
  {Gepr\"ags}}, \bibinfo {author} {\bibfnamefont {Rudolf}\ \bibnamefont
  {Gross}}, \bibinfo {author} {\bibfnamefont {Hans}\ \bibnamefont {Huebl}}, \
  and\ \bibinfo {author} {\bibfnamefont {Matthias}\ \bibnamefont {Althammer}},\
  }\href@noop {} {\enquote {\bibinfo {title} {Observation of antiferromagnetic
  magnon pseudospin dynamics and the hanle effect},}\ } (\bibinfo {year}
  {2020}),\ \bibinfo {note} {\mbox{P}hys. Rev. Lett. (to be published)},\
  \Eprint {http://arxiv.org/abs/2008.00440} {arXiv:2008.00440
  [cond-mat.mtrl-sci]} \BibitemShut {NoStop}%
\bibitem [{\citenamefont {Cheng}\ \emph {et~al.}(2016)\citenamefont {Cheng},
  \citenamefont {Daniels}, \citenamefont {Zhu},\ and\ \citenamefont
  {Xiao}}]{Cheng2016}%
  \BibitemOpen
  \bibfield  {author} {\bibinfo {author} {\bibfnamefont {Ran}\ \bibnamefont
  {Cheng}}, \bibinfo {author} {\bibfnamefont {Matthew~W.}\ \bibnamefont
  {Daniels}}, \bibinfo {author} {\bibfnamefont {Jian-Gang}\ \bibnamefont
  {Zhu}}, \ and\ \bibinfo {author} {\bibfnamefont {Di}~\bibnamefont {Xiao}},\
  }\bibfield  {title} {\enquote {\bibinfo {title} {Antiferromagnetic spin wave
  field-effect transistor},}\ }\href {\doibase 10.1038/srep24223} {\bibfield
  {journal} {\bibinfo  {journal} {Scientific Reports}\ }\textbf {\bibinfo
  {volume} {6}},\ \bibinfo {pages} {24223} (\bibinfo {year}
  {2016})}\BibitemShut {NoStop}%
\bibitem [{\citenamefont {Shen}(2020)}]{Shen2020}%
  \BibitemOpen
  \bibfield  {author} {\bibinfo {author} {\bibfnamefont {Ka}~\bibnamefont
  {Shen}},\ }\bibfield  {title} {\enquote {\bibinfo {title} {Magnon spin
  relaxation and spin hall effect due to the dipolar interaction in
  antiferromagnetic insulators},}\ }\href {\doibase
  10.1103/physrevlett.124.077201} {\bibfield  {journal} {\bibinfo  {journal}
  {Physical Review Letters}\ }\textbf {\bibinfo {volume} {124}},\ \bibinfo
  {pages} {077201} (\bibinfo {year} {2020})}\BibitemShut {NoStop}%
\bibitem [{\citenamefont {Daniels}\ \emph {et~al.}(2018)\citenamefont
  {Daniels}, \citenamefont {Cheng}, \citenamefont {Yu}, \citenamefont {Xiao},\
  and\ \citenamefont {Xiao}}]{Daniels2018}%
  \BibitemOpen
  \bibfield  {author} {\bibinfo {author} {\bibfnamefont {Matthew~W.}\
  \bibnamefont {Daniels}}, \bibinfo {author} {\bibfnamefont {Ran}\ \bibnamefont
  {Cheng}}, \bibinfo {author} {\bibfnamefont {Weichao}\ \bibnamefont {Yu}},
  \bibinfo {author} {\bibfnamefont {Jiang}\ \bibnamefont {Xiao}}, \ and\
  \bibinfo {author} {\bibfnamefont {Di}~\bibnamefont {Xiao}},\ }\bibfield
  {title} {\enquote {\bibinfo {title} {Nonabelian magnonics in
  antiferromagnets},}\ }\href {\doibase 10.1103/PhysRevB.98.134450} {\bibfield
  {journal} {\bibinfo  {journal} {Phys. Rev. B}\ }\textbf {\bibinfo {volume}
  {98}},\ \bibinfo {pages} {134450} (\bibinfo {year} {2018})}\BibitemShut
  {NoStop}%
\bibitem [{\citenamefont {Kawano}\ and\ \citenamefont
  {Hotta}(2019)}]{Kawano2019}%
  \BibitemOpen
  \bibfield  {author} {\bibinfo {author} {\bibfnamefont {Masataka}\
  \bibnamefont {Kawano}}\ and\ \bibinfo {author} {\bibfnamefont {Chisa}\
  \bibnamefont {Hotta}},\ }\bibfield  {title} {\enquote {\bibinfo {title}
  {Thermal hall effect and topological edge states in a square-lattice
  antiferromagnet},}\ }\href {\doibase 10.1103/PhysRevB.99.054422} {\bibfield
  {journal} {\bibinfo  {journal} {Phys. Rev. B}\ }\textbf {\bibinfo {volume}
  {99}},\ \bibinfo {pages} {054422} (\bibinfo {year} {2019})}\BibitemShut
  {NoStop}%
\bibitem [{\citenamefont {Kawano}\ \emph {et~al.}(2019)\citenamefont {Kawano},
  \citenamefont {Onose},\ and\ \citenamefont {Hotta}}]{Kawano2019B}%
  \BibitemOpen
  \bibfield  {author} {\bibinfo {author} {\bibfnamefont {Masataka}\
  \bibnamefont {Kawano}}, \bibinfo {author} {\bibfnamefont {Yoshinori}\
  \bibnamefont {Onose}}, \ and\ \bibinfo {author} {\bibfnamefont {Chisa}\
  \bibnamefont {Hotta}},\ }\bibfield  {title} {\enquote {\bibinfo {title}
  {Designing rashba–dresselhaus effect in magnetic insulators},}\ }\href
  {\doibase 10.1038/s42005-019-0128-6} {\bibfield  {journal} {\bibinfo
  {journal} {Communications Physics}\ }\textbf {\bibinfo {volume} {2}},\
  \bibinfo {pages} {27} (\bibinfo {year} {2019})}\BibitemShut {NoStop}%
\bibitem [{\citenamefont {Hasan}\ and\ \citenamefont {Kane}(2010)}]{Hasan2010}%
  \BibitemOpen
  \bibfield  {author} {\bibinfo {author} {\bibfnamefont {M.~Z.}\ \bibnamefont
  {Hasan}}\ and\ \bibinfo {author} {\bibfnamefont {C.~L.}\ \bibnamefont
  {Kane}},\ }\bibfield  {title} {\enquote {\bibinfo {title} {Colloquium:
  Topological insulators},}\ }\href {\doibase 10.1103/revmodphys.82.3045}
  {\bibfield  {journal} {\bibinfo  {journal} {Reviews of Modern Physics}\
  }\textbf {\bibinfo {volume} {82}},\ \bibinfo {pages} {3045--3067} (\bibinfo
  {year} {2010})}\BibitemShut {NoStop}%
\bibitem [{\citenamefont {Tokura}\ \emph {et~al.}(2019)\citenamefont {Tokura},
  \citenamefont {Yasuda},\ and\ \citenamefont {Tsukazaki}}]{Tokura2019}%
  \BibitemOpen
  \bibfield  {author} {\bibinfo {author} {\bibfnamefont {Yoshinori}\
  \bibnamefont {Tokura}}, \bibinfo {author} {\bibfnamefont {Kenji}\
  \bibnamefont {Yasuda}}, \ and\ \bibinfo {author} {\bibfnamefont {Atsushi}\
  \bibnamefont {Tsukazaki}},\ }\bibfield  {title} {\enquote {\bibinfo {title}
  {Magnetic topological insulators},}\ }\href {\doibase
  10.1038/s42254-018-0011-5} {\bibfield  {journal} {\bibinfo  {journal} {Nature
  Reviews Physics}\ }\textbf {\bibinfo {volume} {1}},\ \bibinfo {pages}
  {126--143} (\bibinfo {year} {2019})}\BibitemShut {NoStop}%
\bibitem [{\citenamefont {Liu}\ \emph {et~al.}(2020)\citenamefont {Liu},
  \citenamefont {Wang},\ and\ \citenamefont {Shen}}]{Liu2020}%
  \BibitemOpen
  \bibfield  {author} {\bibinfo {author} {\bibfnamefont {Jie}\ \bibnamefont
  {Liu}}, \bibinfo {author} {\bibfnamefont {Lin}\ \bibnamefont {Wang}}, \ and\
  \bibinfo {author} {\bibfnamefont {Ka}~\bibnamefont {Shen}},\ }\bibfield
  {title} {\enquote {\bibinfo {title} {Dipolar spin waves in uniaxial easy-axis
  antiferromagnets: A natural topological nodal-line semimetal},}\ }\href
  {\doibase 10.1103/PhysRevResearch.2.023282} {\bibfield  {journal} {\bibinfo
  {journal} {Phys. Rev. Research}\ }\textbf {\bibinfo {volume} {2}},\ \bibinfo
  {pages} {023282} (\bibinfo {year} {2020})}\BibitemShut {NoStop}%
\bibitem [{\citenamefont {Fabian}\ \emph {et~al.}(2007)\citenamefont {Fabian},
  \citenamefont {Matos-Abiague}, \citenamefont {Ertler}, \citenamefont
  {Stano},\ and\ \citenamefont {Zutic}}]{Fabian2007}%
  \BibitemOpen
  \bibfield  {author} {\bibinfo {author} {\bibfnamefont {Jaroslav}\
  \bibnamefont {Fabian}}, \bibinfo {author} {\bibfnamefont {Alex}\ \bibnamefont
  {Matos-Abiague}}, \bibinfo {author} {\bibfnamefont {Christian}\ \bibnamefont
  {Ertler}}, \bibinfo {author} {\bibfnamefont {Peter}\ \bibnamefont {Stano}}, \
  and\ \bibinfo {author} {\bibfnamefont {Igor}\ \bibnamefont {Zutic}},\
  }\bibfield  {title} {\enquote {\bibinfo {title} {Semiconductor
  spintronics},}\ }\href
  {http://www.physics.sk/aps/pub.php?y=2007&pub=aps-07-04} {\bibfield
  {journal} {\bibinfo  {journal} {Acta Physica Slovaca}\ }\textbf {\bibinfo
  {volume} {57}},\ \bibinfo {pages} {565} (\bibinfo {year} {2007})}\BibitemShut
  {NoStop}%
\bibitem [{\citenamefont {Wu}\ \emph {et~al.}(2010)\citenamefont {Wu},
  \citenamefont {Jiang},\ and\ \citenamefont {Weng}}]{Wu2010}%
  \BibitemOpen
  \bibfield  {author} {\bibinfo {author} {\bibfnamefont {M.W.}\ \bibnamefont
  {Wu}}, \bibinfo {author} {\bibfnamefont {J.H.}\ \bibnamefont {Jiang}}, \ and\
  \bibinfo {author} {\bibfnamefont {M.Q.}\ \bibnamefont {Weng}},\ }\bibfield
  {title} {\enquote {\bibinfo {title} {Spin dynamics in semiconductors},}\
  }\href {\doibase https://doi.org/10.1016/j.physrep.2010.04.002} {\bibfield
  {journal} {\bibinfo  {journal} {Physics Reports}\ }\textbf {\bibinfo {volume}
  {493}},\ \bibinfo {pages} {61 -- 236} (\bibinfo {year} {2010})}\BibitemShut
  {NoStop}%
\bibitem [{\citenamefont {Ross}\ \emph
  {et~al.}(2020{\natexlab{b}})\citenamefont {Ross}, \citenamefont {Lebrun},
  \citenamefont {Baldrati}, \citenamefont {Kamra}, \citenamefont {Gomonay},
  \citenamefont {Ding}, \citenamefont {Schreiber}, \citenamefont {Backes},
  \citenamefont {Maccherozzi}, \citenamefont {Grave}, \citenamefont
  {Rothschild}, \citenamefont {Sinova},\ and\ \citenamefont
  {Kl\"aui}}]{Ross2020}%
  \BibitemOpen
  \bibfield  {author} {\bibinfo {author} {\bibfnamefont {Andrew}\ \bibnamefont
  {Ross}}, \bibinfo {author} {\bibfnamefont {Romain}\ \bibnamefont {Lebrun}},
  \bibinfo {author} {\bibfnamefont {Lorenzo}\ \bibnamefont {Baldrati}},
  \bibinfo {author} {\bibfnamefont {Akashdeep}\ \bibnamefont {Kamra}}, \bibinfo
  {author} {\bibfnamefont {Olena}\ \bibnamefont {Gomonay}}, \bibinfo {author}
  {\bibfnamefont {Shilei}\ \bibnamefont {Ding}}, \bibinfo {author}
  {\bibfnamefont {Felix}\ \bibnamefont {Schreiber}}, \bibinfo {author}
  {\bibfnamefont {Dirk}\ \bibnamefont {Backes}}, \bibinfo {author}
  {\bibfnamefont {Francesco}\ \bibnamefont {Maccherozzi}}, \bibinfo {author}
  {\bibfnamefont {Daniel~A.}\ \bibnamefont {Grave}}, \bibinfo {author}
  {\bibfnamefont {Avner}\ \bibnamefont {Rothschild}}, \bibinfo {author}
  {\bibfnamefont {Jairo}\ \bibnamefont {Sinova}}, \ and\ \bibinfo {author}
  {\bibfnamefont {Mathias}\ \bibnamefont {Kl\"aui}},\ }\href@noop {} {\enquote
  {\bibinfo {title} {An insulating doped antiferromagnet with low magnetic
  symmetry as a room temperature spin conduit},}\ } (\bibinfo {year}
  {2020}{\natexlab{b}}),\ \Eprint {http://arxiv.org/abs/2011.09755}
  {arXiv:2011.09755 [cond-mat.mes-hall]} \BibitemShut {NoStop}%
\bibitem [{\citenamefont {Lan}\ \emph {et~al.}(2017)\citenamefont {Lan},
  \citenamefont {Yu},\ and\ \citenamefont {Xiao}}]{Lan2017}%
  \BibitemOpen
  \bibfield  {author} {\bibinfo {author} {\bibfnamefont {Jin}\ \bibnamefont
  {Lan}}, \bibinfo {author} {\bibfnamefont {Weichao}\ \bibnamefont {Yu}}, \
  and\ \bibinfo {author} {\bibfnamefont {Jiang}\ \bibnamefont {Xiao}},\
  }\bibfield  {title} {\enquote {\bibinfo {title} {Antiferromagnetic domain
  wall as spin wave polarizer and retarder},}\ }\href {\doibase
  10.1038/s41467-017-00265-5} {\bibfield  {journal} {\bibinfo  {journal}
  {Nature Communications}\ }\textbf {\bibinfo {volume} {8}},\ \bibinfo {pages}
  {2041--1723} (\bibinfo {year} {2017})}\BibitemShut {NoStop}%
\bibitem [{\citenamefont {Flebus}(2019)}]{Flebus2019}%
  \BibitemOpen
  \bibfield  {author} {\bibinfo {author} {\bibfnamefont {Benedetta}\
  \bibnamefont {Flebus}},\ }\bibfield  {title} {\enquote {\bibinfo {title}
  {Chemical potential of an antiferromagnetic magnon gas},}\ }\href {\doibase
  10.1103/PhysRevB.100.064410} {\bibfield  {journal} {\bibinfo  {journal}
  {Phys. Rev. B}\ }\textbf {\bibinfo {volume} {100}},\ \bibinfo {pages}
  {064410} (\bibinfo {year} {2019})}\BibitemShut {NoStop}%
\bibitem [{\citenamefont {Keffer}\ and\ \citenamefont
  {Loudon}(1961)}]{Keffer1961}%
  \BibitemOpen
  \bibfield  {author} {\bibinfo {author} {\bibfnamefont {F.}~\bibnamefont
  {Keffer}}\ and\ \bibinfo {author} {\bibfnamefont {R.}~\bibnamefont
  {Loudon}},\ }\bibfield  {title} {\enquote {\bibinfo {title} {Simple physical
  theory of spin wave interactions},}\ }\href {\doibase 10.1063/1.2000447}
  {\bibfield  {journal} {\bibinfo  {journal} {Journal of Applied Physics}\
  }\textbf {\bibinfo {volume} {32}},\ \bibinfo {pages} {S2--S7} (\bibinfo
  {year} {1961})}\BibitemShut {NoStop}%
\bibitem [{\citenamefont {Bloch}(1962)}]{Bloch1962}%
  \BibitemOpen
  \bibfield  {author} {\bibinfo {author} {\bibfnamefont {Micheline}\
  \bibnamefont {Bloch}},\ }\bibfield  {title} {\enquote {\bibinfo {title}
  {Magnon renormalization in ferromagnets near the curie point},}\ }\href
  {\doibase 10.1103/PhysRevLett.9.286} {\bibfield  {journal} {\bibinfo
  {journal} {Phys. Rev. Lett.}\ }\textbf {\bibinfo {volume} {9}},\ \bibinfo
  {pages} {286--287} (\bibinfo {year} {1962})}\BibitemShut {NoStop}%
\bibitem [{\citenamefont {Kamra}\ \emph {et~al.}(2019)\citenamefont {Kamra},
  \citenamefont {Thingstad}, \citenamefont {Rastelli}, \citenamefont {Duine},
  \citenamefont {Brataas}, \citenamefont {Belzig},\ and\ \citenamefont
  {Sudb\o{}}}]{Kamra2019}%
  \BibitemOpen
  \bibfield  {author} {\bibinfo {author} {\bibfnamefont {Akashdeep}\
  \bibnamefont {Kamra}}, \bibinfo {author} {\bibfnamefont {Even}\ \bibnamefont
  {Thingstad}}, \bibinfo {author} {\bibfnamefont {Gianluca}\ \bibnamefont
  {Rastelli}}, \bibinfo {author} {\bibfnamefont {Rembert~A.}\ \bibnamefont
  {Duine}}, \bibinfo {author} {\bibfnamefont {Arne}\ \bibnamefont {Brataas}},
  \bibinfo {author} {\bibfnamefont {Wolfgang}\ \bibnamefont {Belzig}}, \ and\
  \bibinfo {author} {\bibfnamefont {Asle}\ \bibnamefont {Sudb\o{}}},\
  }\bibfield  {title} {\enquote {\bibinfo {title} {Antiferromagnetic magnons as
  highly squeezed fock states underlying quantum correlations},}\ }\href
  {\doibase 10.1103/PhysRevB.100.174407} {\bibfield  {journal} {\bibinfo
  {journal} {Phys. Rev. B}\ }\textbf {\bibinfo {volume} {100}},\ \bibinfo
  {pages} {174407} (\bibinfo {year} {2019})}\BibitemShut {NoStop}%
\bibitem [{\citenamefont {Liensberger}\ \emph {et~al.}(2019)\citenamefont
  {Liensberger}, \citenamefont {Kamra}, \citenamefont {Maier-Flaig},
  \citenamefont {Gepr\"ags}, \citenamefont {Erb}, \citenamefont {Goennenwein},
  \citenamefont {Gross}, \citenamefont {Belzig}, \citenamefont {Huebl},\ and\
  \citenamefont {Weiler}}]{Liensberger2019}%
  \BibitemOpen
  \bibfield  {author} {\bibinfo {author} {\bibfnamefont {Lukas}\ \bibnamefont
  {Liensberger}}, \bibinfo {author} {\bibfnamefont {Akashdeep}\ \bibnamefont
  {Kamra}}, \bibinfo {author} {\bibfnamefont {Hannes}\ \bibnamefont
  {Maier-Flaig}}, \bibinfo {author} {\bibfnamefont {Stephan}\ \bibnamefont
  {Gepr\"ags}}, \bibinfo {author} {\bibfnamefont {Andreas}\ \bibnamefont
  {Erb}}, \bibinfo {author} {\bibfnamefont {Sebastian T.~B.}\ \bibnamefont
  {Goennenwein}}, \bibinfo {author} {\bibfnamefont {Rudolf}\ \bibnamefont
  {Gross}}, \bibinfo {author} {\bibfnamefont {Wolfgang}\ \bibnamefont
  {Belzig}}, \bibinfo {author} {\bibfnamefont {Hans}\ \bibnamefont {Huebl}}, \
  and\ \bibinfo {author} {\bibfnamefont {Mathias}\ \bibnamefont {Weiler}},\
  }\bibfield  {title} {\enquote {\bibinfo {title} {Exchange-enhanced
  ultrastrong magnon-magnon coupling in a compensated ferrimagnet},}\ }\href
  {\doibase 10.1103/PhysRevLett.123.117204} {\bibfield  {journal} {\bibinfo
  {journal} {Phys. Rev. Lett.}\ }\textbf {\bibinfo {volume} {123}},\ \bibinfo
  {pages} {117204} (\bibinfo {year} {2019})}\BibitemShut {NoStop}%
\bibitem [{Note2()}]{Note2}%
  \BibitemOpen
  \bibinfo {note} {As per our chosen convention, the sense of precession for
  magnon pseudospin about pseudofield is opposite to that of electron spin
  about a magnetic field due to the negative gyromagnetic ratio of the
  latter.}\BibitemShut {Stop}%
\bibitem [{Note3()}]{Note3}%
  \BibitemOpen
  \bibinfo {note} {In this representation, an overall phase factor in both the
  eigenvectors has been disregarded.}\BibitemShut {Stop}%
\bibitem [{\citenamefont {Jones}\ \emph {et~al.}(2016)\citenamefont {Jones},
  \citenamefont {D’Addario}, \citenamefont {Rojec}, \citenamefont {Milione},\
  and\ \citenamefont {Galvez}}]{Jones2016}%
  \BibitemOpen
  \bibfield  {author} {\bibinfo {author} {\bibfnamefont {Joshua~A.}\
  \bibnamefont {Jones}}, \bibinfo {author} {\bibfnamefont {Anthony~J.}\
  \bibnamefont {D’Addario}}, \bibinfo {author} {\bibfnamefont {Brett~L.}\
  \bibnamefont {Rojec}}, \bibinfo {author} {\bibfnamefont {G.}~\bibnamefont
  {Milione}}, \ and\ \bibinfo {author} {\bibfnamefont {Enrique~J.}\
  \bibnamefont {Galvez}},\ }\bibfield  {title} {\enquote {\bibinfo {title} {The
  poincaré-sphere approach to polarization: Formalism and new labs with
  poincaré beams},}\ }\href {\doibase 10.1119/1.4960468} {\bibfield  {journal}
  {\bibinfo  {journal} {American Journal of Physics}\ }\textbf {\bibinfo
  {volume} {84}},\ \bibinfo {pages} {822--835} (\bibinfo {year}
  {2016})}\BibitemShut {NoStop}%
\bibitem [{\citenamefont {Kikkawa}\ and\ \citenamefont
  {Awschalom}(1999)}]{Kikkawa1999}%
  \BibitemOpen
  \bibfield  {author} {\bibinfo {author} {\bibfnamefont {J.~M.}\ \bibnamefont
  {Kikkawa}}\ and\ \bibinfo {author} {\bibfnamefont {D.~D.}\ \bibnamefont
  {Awschalom}},\ }\bibfield  {title} {\enquote {\bibinfo {title} {Lateral drag
  of spin coherence in gallium arsenide},}\ }\href {\doibase 10.1038/16420}
  {\bibfield  {journal} {\bibinfo  {journal} {Nature}\ }\textbf {\bibinfo
  {volume} {397}},\ \bibinfo {pages} {139--141} (\bibinfo {year}
  {1999})}\BibitemShut {NoStop}%
\bibitem [{\citenamefont {Jedema}\ \emph {et~al.}(2002)\citenamefont {Jedema},
  \citenamefont {Heersche}, \citenamefont {Filip}, \citenamefont {Baselmans},\
  and\ \citenamefont {van Wees}}]{Jedema2002}%
  \BibitemOpen
  \bibfield  {author} {\bibinfo {author} {\bibfnamefont {F.~J.}\ \bibnamefont
  {Jedema}}, \bibinfo {author} {\bibfnamefont {H.~B.}\ \bibnamefont
  {Heersche}}, \bibinfo {author} {\bibfnamefont {A.~T.}\ \bibnamefont {Filip}},
  \bibinfo {author} {\bibfnamefont {J.~J.~A.}\ \bibnamefont {Baselmans}}, \
  and\ \bibinfo {author} {\bibfnamefont {B.~J.}\ \bibnamefont {van Wees}},\
  }\bibfield  {title} {\enquote {\bibinfo {title} {Electrical detection of spin
  precession in a metallic mesoscopic spin valve},}\ }\href {\doibase
  10.1038/416713a} {\bibfield  {journal} {\bibinfo  {journal} {Nature}\
  }\textbf {\bibinfo {volume} {416}},\ \bibinfo {pages} {713--716} (\bibinfo
  {year} {2002})}\BibitemShut {NoStop}%
\bibitem [{\citenamefont {Cornelissen}\ \emph
  {et~al.}(2016{\natexlab{b}})\citenamefont {Cornelissen}, \citenamefont
  {Shan},\ and\ \citenamefont {van Wees}}]{CornelissenTemp}%
  \BibitemOpen
  \bibfield  {author} {\bibinfo {author} {\bibfnamefont {L.~J.}\ \bibnamefont
  {Cornelissen}}, \bibinfo {author} {\bibfnamefont {J.}~\bibnamefont {Shan}}, \
  and\ \bibinfo {author} {\bibfnamefont {B.~J.}\ \bibnamefont {van Wees}},\
  }\bibfield  {title} {\enquote {\bibinfo {title} {Temperature dependence of
  the magnon spin diffusion length and magnon spin conductivity in the magnetic
  insulator yttrium iron garnet},}\ }\href {\doibase
  10.1103/physrevb.94.180402} {\bibfield  {journal} {\bibinfo  {journal}
  {Physical Review B}\ }\textbf {\bibinfo {volume} {94}},\ \bibinfo {pages}
  {180402} (\bibinfo {year} {2016}{\natexlab{b}})}\BibitemShut {NoStop}%
\bibitem [{\citenamefont {Shan}\ \emph {et~al.}(2017)\citenamefont {Shan},
  \citenamefont {Bougiatioti}, \citenamefont {Liang}, \citenamefont {Reiss},
  \citenamefont {Kuschel},\ and\ \citenamefont {van Wees}}]{Shan2017}%
  \BibitemOpen
  \bibfield  {author} {\bibinfo {author} {\bibfnamefont {J.}~\bibnamefont
  {Shan}}, \bibinfo {author} {\bibfnamefont {P.}~\bibnamefont {Bougiatioti}},
  \bibinfo {author} {\bibfnamefont {L.}~\bibnamefont {Liang}}, \bibinfo
  {author} {\bibfnamefont {G.}~\bibnamefont {Reiss}}, \bibinfo {author}
  {\bibfnamefont {T.}~\bibnamefont {Kuschel}}, \ and\ \bibinfo {author}
  {\bibfnamefont {B.~J.}\ \bibnamefont {van Wees}},\ }\bibfield  {title}
  {\enquote {\bibinfo {title} {Nonlocal magnon spin transport in {NiFe}2o4 thin
  films},}\ }\href {\doibase 10.1063/1.4979408} {\bibfield  {journal} {\bibinfo
   {journal} {Applied Physics Letters}\ }\textbf {\bibinfo {volume} {110}},\
  \bibinfo {pages} {132406} (\bibinfo {year} {2017})}\BibitemShut {NoStop}%
\bibitem [{\citenamefont {Aspelmeyer}\ \emph {et~al.}(2014)\citenamefont
  {Aspelmeyer}, \citenamefont {Kippenberg},\ and\ \citenamefont
  {Marquardt}}]{Aspelmeyer2014}%
  \BibitemOpen
  \bibfield  {author} {\bibinfo {author} {\bibfnamefont {Markus}\ \bibnamefont
  {Aspelmeyer}}, \bibinfo {author} {\bibfnamefont {Tobias~J.}\ \bibnamefont
  {Kippenberg}}, \ and\ \bibinfo {author} {\bibfnamefont {Florian}\
  \bibnamefont {Marquardt}},\ }\bibfield  {title} {\enquote {\bibinfo {title}
  {Cavity optomechanics},}\ }\href {\doibase 10.1103/RevModPhys.86.1391}
  {\bibfield  {journal} {\bibinfo  {journal} {Rev. Mod. Phys.}\ }\textbf
  {\bibinfo {volume} {86}},\ \bibinfo {pages} {1391--1452} (\bibinfo {year}
  {2014})}\BibitemShut {NoStop}%
\bibitem [{\citenamefont {MacNeill}\ \emph {et~al.}(2019)\citenamefont
  {MacNeill}, \citenamefont {Hou}, \citenamefont {Klein}, \citenamefont
  {Zhang}, \citenamefont {Jarillo-Herrero},\ and\ \citenamefont
  {Liu}}]{MacNeill2019}%
  \BibitemOpen
  \bibfield  {author} {\bibinfo {author} {\bibfnamefont {David}\ \bibnamefont
  {MacNeill}}, \bibinfo {author} {\bibfnamefont {Justin~T.}\ \bibnamefont
  {Hou}}, \bibinfo {author} {\bibfnamefont {Dahlia~R.}\ \bibnamefont {Klein}},
  \bibinfo {author} {\bibfnamefont {Pengxiang}\ \bibnamefont {Zhang}}, \bibinfo
  {author} {\bibfnamefont {Pablo}\ \bibnamefont {Jarillo-Herrero}}, \ and\
  \bibinfo {author} {\bibfnamefont {Luqiao}\ \bibnamefont {Liu}},\ }\bibfield
  {title} {\enquote {\bibinfo {title} {Gigahertz frequency antiferromagnetic
  resonance and strong magnon-magnon coupling in the layered crystal
  ${\mathrm{crcl}}_{3}$},}\ }\href {\doibase 10.1103/PhysRevLett.123.047204}
  {\bibfield  {journal} {\bibinfo  {journal} {Phys. Rev. Lett.}\ }\textbf
  {\bibinfo {volume} {123}},\ \bibinfo {pages} {047204} (\bibinfo {year}
  {2019})}\BibitemShut {NoStop}%
\bibitem [{\citenamefont {Viola~Kusminskiy}\ \emph {et~al.}(2016)\citenamefont
  {Viola~Kusminskiy}, \citenamefont {Tang},\ and\ \citenamefont
  {Marquardt}}]{Kusminskiy2016}%
  \BibitemOpen
  \bibfield  {author} {\bibinfo {author} {\bibfnamefont {Silvia}\ \bibnamefont
  {Viola~Kusminskiy}}, \bibinfo {author} {\bibfnamefont {Hong~X.}\ \bibnamefont
  {Tang}}, \ and\ \bibinfo {author} {\bibfnamefont {Florian}\ \bibnamefont
  {Marquardt}},\ }\bibfield  {title} {\enquote {\bibinfo {title} {Coupled
  spin-light dynamics in cavity optomagnonics},}\ }\href {\doibase
  10.1103/PhysRevA.94.033821} {\bibfield  {journal} {\bibinfo  {journal} {Phys.
  Rev. A}\ }\textbf {\bibinfo {volume} {94}},\ \bibinfo {pages} {033821}
  (\bibinfo {year} {2016})}\BibitemShut {NoStop}%
\bibitem [{\citenamefont {Harder}\ and\ \citenamefont {Hu}(2018)}]{Harder2018}%
  \BibitemOpen
  \bibfield  {author} {\bibinfo {author} {\bibfnamefont {Michael}\ \bibnamefont
  {Harder}}\ and\ \bibinfo {author} {\bibfnamefont {Can-Ming}\ \bibnamefont
  {Hu}},\ }\bibfield  {title} {\enquote {\bibinfo {title} {{Cavity Spintronics:
  An Early Review of Recent Progress in the Study of Magnon–Photon Level
  Repulsion}},}\ }in\ \href {\doibase 10.1016/bs.ssp.2018.08.001} {\emph
  {\bibinfo {booktitle} {Solid State Physics 69}}},\ \bibinfo {editor} {edited
  by\ \bibinfo {editor} {\bibfnamefont {Robert~E.}\ \bibnamefont {Camley}}\
  and\ \bibinfo {editor} {\bibfnamefont {Robert~L.}\ \bibnamefont {Stamps}}}\
  (\bibinfo  {publisher} {Academic Press},\ \bibinfo {address} {Cambridge},\
  \bibinfo {year} {2018})\ pp.\ \bibinfo {pages} {47--121}\BibitemShut
  {NoStop}%
\bibitem [{\citenamefont {Huebl}\ \emph {et~al.}(2013)\citenamefont {Huebl},
  \citenamefont {Zollitsch}, \citenamefont {Lotze}, \citenamefont {Hocke},
  \citenamefont {Greifenstein}, \citenamefont {Marx}, \citenamefont {Gross},\
  and\ \citenamefont {Goennenwein}}]{Huebl2013}%
  \BibitemOpen
  \bibfield  {author} {\bibinfo {author} {\bibfnamefont {Hans}\ \bibnamefont
  {Huebl}}, \bibinfo {author} {\bibfnamefont {Christoph~W.}\ \bibnamefont
  {Zollitsch}}, \bibinfo {author} {\bibfnamefont {Johannes}\ \bibnamefont
  {Lotze}}, \bibinfo {author} {\bibfnamefont {Fredrik}\ \bibnamefont {Hocke}},
  \bibinfo {author} {\bibfnamefont {Moritz}\ \bibnamefont {Greifenstein}},
  \bibinfo {author} {\bibfnamefont {Achim}\ \bibnamefont {Marx}}, \bibinfo
  {author} {\bibfnamefont {Rudolf}\ \bibnamefont {Gross}}, \ and\ \bibinfo
  {author} {\bibfnamefont {Sebastian T.~B.}\ \bibnamefont {Goennenwein}},\
  }\bibfield  {title} {\enquote {\bibinfo {title} {{High Cooperativity in
  Coupled Microwave Resonator Ferrimagnetic Insulator Hybrids}},}\ }\href
  {\doibase 10.1103/PhysRevLett.111.127003} {\bibfield  {journal} {\bibinfo
  {journal} {Physical Review Letters}\ }\textbf {\bibinfo {volume} {111}},\
  \bibinfo {pages} {127003} (\bibinfo {year} {2013})}\BibitemShut {NoStop}%
\bibitem [{\citenamefont {Holstein}\ and\ \citenamefont
  {Primakoff}(1940)}]{Holstein1940}%
  \BibitemOpen
  \bibfield  {author} {\bibinfo {author} {\bibfnamefont {T.}~\bibnamefont
  {Holstein}}\ and\ \bibinfo {author} {\bibfnamefont {H.}~\bibnamefont
  {Primakoff}},\ }\bibfield  {title} {\enquote {\bibinfo {title} {Field
  dependence of the intrinsic domain magnetization of a ferromagnet},}\ }\href
  {\doibase 10.1103/PhysRev.58.1098} {\bibfield  {journal} {\bibinfo  {journal}
  {Phys. Rev.}\ }\textbf {\bibinfo {volume} {58}},\ \bibinfo {pages}
  {1098--1113} (\bibinfo {year} {1940})}\BibitemShut {NoStop}%
\bibitem [{\citenamefont {Akhiezer}\ \emph {et~al.}(1968)\citenamefont
  {Akhiezer}, \citenamefont {Bar'iakhtar},\ and\ \citenamefont
  {Peletminski}}]{Akhiezer1968}%
  \BibitemOpen
  \bibfield  {author} {\bibinfo {author} {\bibfnamefont {A.I.}\ \bibnamefont
  {Akhiezer}}, \bibinfo {author} {\bibfnamefont {V.G.}\ \bibnamefont
  {Bar'iakhtar}}, \ and\ \bibinfo {author} {\bibfnamefont {S.V.}\ \bibnamefont
  {Peletminski}},\ }\href {http://books.google.nl/books?id=GpA6AAAAMAAJ} {\emph
  {\bibinfo {title} {Spin waves}}}\ (\bibinfo  {publisher} {North-Holland
  Publishing Company, Amsterdam},\ \bibinfo {year} {1968})\BibitemShut
  {NoStop}%
\end{thebibliography}%

\end{document}